\documentclass[longauth]{aa} 

\usepackage{graphicx}
\usepackage{graphicx,xcolor}
\usepackage{txfonts}
\usepackage{multirow}
\usepackage{lscape}
\usepackage{longtable}
\usepackage{comment}
\usepackage{placeins}
\usepackage[breaklinks=true]{hyperref}

    \hypersetup{
        colorlinks=true,
        linkcolor=blue,
        filecolor=magenta,      
        urlcolor=blue,
        citecolor=blue
    }
\usepackage{tabularx}
\usepackage{natbib}
\usepackage[export]{adjustbox}
\usepackage{soul}

\newcommand{\HI}{\textrm{H~{\textsc{i}}}}

\newcommand{\molh}{H$_2$\xspace}

\newcommand{\Tb}{atomic PDR\xspace}

\usepackage{color}
\usepackage{soul}

\usepackage{xcolor}

\definecolor{cbpurple}{rgb}{0.47, 0.37, 0.94}

\begin{document}

   \title{PDRs4All \textsc{V}: Modelling the dust evolution across the illuminated edge of the Orion Bar\thanks{Based on the ERS project PDRs4All \#1288 \emph{"Radiative Feedback from Massive Stars as Traced by Multiband Imaging and Spectroscopic Mosaics"} observations obtained with JWST instruments (\url{https://www.stsci.edu/jwst/}), a NASA/ESA/CSA science mission with instruments and contributions directly funded by ESA Member States, NASA, and Canada}
}

\titlerunning{Modelling the dust evolution across the illuminated egde of the Orion Bar}

\author{M. Elyajouri\inst{1} \and
        N. Ysard\inst{2,1} \and
        A. Abergel\inst{1} \and
        E. Habart\inst{1} \and
        L. Verstraete\inst{1} \and
        A. Jones\inst{1} \and
        M. Juvela\inst{3} \and
        T. Schirmer\inst{4} \and
        R. Meshaka\inst{1,6} \and
        E. Dartois\inst{5} \and
        J. Lebourlot\inst{6} \and
        G. Rouill\'{e}\inst{7} \and
        T. Onaka\inst{8} \and
        E. Peeters\inst{9,10,11}\and
        O. Bern\'{e}\inst{2}\and
        F. Alarc\'on \inst{12} \and
        J. Bernard-Salas\inst{13,14}\and
        M. Buragohain\inst{15} \and
        J. Cami\inst{8,9,10} \and
        A. Canin \inst{2} \and
        R. Chown \inst{9,10}\and
        K. Demyk \inst{2} \and
        K. Gordon\inst{16,17} \and
        O. Kannavou \inst{1} \and
        M. Kirsanova\inst{18} \and
        S. Madden\inst{19} \and
        R. Paladini\inst{20} \and
        Y. Pendleton\inst{21} \and
        F. Salama\inst{22} \and  
        I. Schroetter \inst{2} \and
        A. Sidhu \inst{9,10} \and
        M. R\"{o}llig\inst{23,24}
        B. Trahin \inst{1} \and
        D. Van De Putte \inst{16} 
}

\institute{
    Institut d'Astrophysique Spatiale, Université Paris-Saclay, CNRS, Orsay, France \label{inst1} \and
    Institut de Recherche en Astrophysique et Planétologie, Toulouse, France \label{inst2}
    \and
    Department of Physics, University of Helsinki, Finland \label{inst3} 
    \and
    Department of Space, Earth and Environment, Chalmers University of Technology, Onsala Space Observatory, Sweden \label{inst4} 
    \and
    Institut des Sciences Moléculaires d'Orsay, CNRS, Université Paris-Saclay, Orsay, France\label{inst5} 
    \and
    Observatoire de Paris, PSL University, Sorbonne Université, LERMA, Paris, France \label{inst6} \and
    Laboratory Astrophysics Group of the Max Planck Institute for Astronomy at the Friedrich Schiller University Jena, Jena, Germany\label{inst7}
    \and
    Department of Astronomy, Graduate School of Science, The University of Tokyo, Tokyo, Japan\label{inst8}
    \and
    Department of Physics \& Astronomy, The University of Western Ontario, London, Canada \label{inst9} \and 
    Institute for Earth and Space Exploration, The University of Western Ontario, London, Canada \label{inst10} \and
    Carl Sagan Center, SETI Institute, Mountain View, CA, USA \label{inst11} \and
    Astronomy Department, University of Maryland, College Park, MD 20742, USA  \label{inst12} \and
    ACRI-ST, Centre d’Etudes et de Recherche de Grasse (CERGA), France \label{inst13} \and
    INCLASS Common Laboratory., Grasse, France \label{inst14} \and
    DST INSPIRE School of Physics, University of Hyderabad, Hyderabad, India \label{inst15} \and
    Space Telescope Science Institute, Baltimore, MD, USA \label{inst16} \and
    Sterrenkundig Observatorium, Universiteit Gent, Gent, Belgium \label{inst17} \and
    Institute of Astronomy, Russian Academy of Sciences, Moscow, Russia \label{inst18} \and
    AIM, CEA, CNRS, Université Paris-Saclay, Université Paris Diderot, Sorbonne Paris Cité, Gif-sur-Yvette, France \label{inst19}  \and
    California Institute of Technology, IPAC, Pasadena, CA, USA \label{inst20} \and
    NASA Ames Research Center, Moffett Field, CA, USA \label{inst21} \and
    Physikalischer Verein - Gesellschaft für Bildung und Wissenschaft, Frankfurt am Main, Germany \label{inst22} \and
    Institut für Angewandte Physik, Goethe-Universität Frankfurt, Frankfurt am Main, Germany \label{inst23} 
}

   \date{Received 29/11/2023; accepted 21/12/2023}

  \abstract
   {Interstellar dust particles, in particular carbonaceous nano-grains (like polycyclic aromatic hydrocarbons, fullerenes, and amorphous hydrogenated carbon), are critical players for the composition, energy budget, and dynamics of the interstellar medium (ISM). The dust properties, specifically the composition and size of dust grains are not static; instead, they exhibit considerable evolution triggered by variations in local physical conditions such as the density and gas temperature within the ISM, as is the case in photon-dominated regions (PDRs). The evolution of dust and its impact on the local physical and chemical conditions is thus a key question for understanding the first stages of star formation. 
   }
   {From the extensive spectral and imaging data of the JWST PDRs4All program, we study the emission of dust grains within the Orion Bar — a well-known, highly far-UV (FUV)-irradiated PDR situated at the intersection between cold, dense molecular clouds, and warm ionized regions. The Orion Bar because of its edge-on geometry provides an exceptional benchmark for characterizing dust evolution and the associated driving processes under varying physical conditions. Our goal is to constrain the local properties of dust by comparing its emission to models. Taking advantage of the recent JWST data, in particular the spectroscopy of dust emission, we identify new constraints on dust and further previous works of dust modelling.
   } 
   {To characterize interstellar dust across the Orion Bar, we follow its emission as traced by JWST NIRCam (at 3.35 and 4.8 $\mu$m) and MIRI (at 7.7, 11.3, 15.0, and 25.5 $\mu$m) broad band images, along with NIRSpec and MRS spectroscopic observations. First, we constrain the minimum size and hydrogen content of carbon nano-grains from a comparison between the observed dust emission spectra and the predictions of the Heterogeneous dust Evolution Model for Interstellar Solids (THEMIS) coupled to the numerical code DustEM. 
   Using this dust model, we then perform 3D radiative transfer simulations of dust emission with the SOC code (Scattering with OpenCL) and compare to data obtained along well chosen profiles across the Orion Bar. 
   }
   {The JWST data allows us, for the first time, to spatially resolve the steep variation of dust emission at the illuminated edge of the Orion Bar PDR. By considering a dust model with carbonaceous nano-grains and submicronic coated silicate grains, we derive unprecedented constraints on the properties of across the Orion Bar. To explain the observed emission profiles with our simulations, we find that the nano-grains must be strongly depleted with an abundance (relative to the gas) 15 times less than in the diffuse ISM. The NIRSpec and MRS spectroscopic observations reveal variations in the hydrogenation of the carbon nano-grains. The lowest hydrogenation levels are found in the vicinity of the illuminating stars suggesting photo-processing while more hydrogenated nano-grains are found in the cold and dense molecular region, potentially indicative of larger grains.
   }
   {}

   \maketitle
\section{Introduction}
Interstellar dust is an essential component of interstellar matter (ISM) and plays a key role in the formation of stars and protoplanetary disks. Ubiquitous in the ISM with a dust-to-gas mass ratio of a percent, dust is formed of tiny solid grains with sizes between a few 0.1 nm and a few 100 nm. Analysis of ISM observations has long shown that dust grains control the transfer of radiation, heat the gas via the photoelectric effect and form efficiently important molecules like H$_2$ on their surface. Dust grains also carry a significant part of the ISM charge and follow magnetic field lines thus coupling gas motions to the magnetic field. Dust grains are therefore a central interface through which redistribution of the ISM energy (in radiation, gas motions or magnetic field) occurs. Because of this, dust observables, and in particular dust emission, is a powerful tracer of the physical conditions in the regions observed (gas density, mass, and intensity of radiation field). Indeed, dust emits in the infrared (IR) to submillimetre wavelength range the energy it has gained from absorption of the local radiation field or from inelastic collisions with gas species, a conversion that allows to probe deep in the clouds because of the smaller extinction at these wavelengths. Yet dust grains are not passive tracers of the ISM because their properties change with the local physical conditions. Grains are eroded by shocks or photo-fragmentation in strongly irradiated regions \citep{pilleri2012,micelotta2010} or grow by coagulation in dense, cold gas \citep{kohler2012,ysard2013,ysard2016} and these processes thus drive the dust evolution. This evolution may change the abundances of the dust populations 
but also the structure (porous, amorphous) and composition of the grains. These latter changes are however more difficult to derive from observations because they require detailed and spectrally extended dust properties for models of dust analogues. Conversely, abundance and size distribution changes of grains are directly reflected in the dust emission. In this context, former work suggested that the main impact of dust evolution is to change the relative abundance of the small (1 nm in size) to large (100 nm in size) grains \citep{arab2012,schirmer2020}. 
In addition, the observed large variation of carbon gas-phase abundance (see references in \cite{compiegne2011}) suggests that this element is efficiently cycled in and out of grains by accretion and destruction processes \citep{Jones2014}. The gas-phase abundance of carbon is thus enhanced by energetic processes (shocks or UV light) which break up dust into smaller carbonaceous grains. This active carbon cycle and extinction observations \citep{parvathi2012} suggest that the small dust population is in fact carbonaceous nano-grains. Proposed to explain the aromatic infrared bands (AIBs) observed between 3 and 20 $\mu$m, these nano-grains must be, at least partially, aromatic and hydrogenated. Plausible interstellar nano-grains range from compact, symmetric species such as Polycyclic Aromatic Hydrocarbons (PAHs) or fullerenes \citep{leger_puget84, allamandola_polycyclic_1985, sellgren2010c60} to more disordered ones such as amorphous hydrogenated carbon or a-C:H \citep{Jones2013} . It must be noted here that models predict that nano-grains (i) dominate the gas photoelectric heating rate \citep{bakesandtielens94,Habart2001,berne2022}, (ii) carry most of the surface for H$_2$ formation 
\citep{bron14,jones2015}, and (iii) are the carriers of the UV dust extinction \citep{weingartner_Draine_01_sizedistrib}. Describing the physical state of the gas that will form stars thus requires knowledge of the properties and abundance variations of nano-grains. 

Ever since the advent of UV and IR data in astronomy, dust models have been designed to explain the main observables (emission and extinction) without evolution. Recent data collected during the ISO, Akari, Spitzer, Planck, and Herschel missions have brought a first look at dust evolution however limited by the low angular resolution \citep[e.g.][]{compiegne2008,arab2012,mori2012,mori2014,pilleri2015,schirmer2022}. 
The advent of JWST data offers new opportunities to study and model dust evolution. In our Galaxy and in external galaxies, interstellar matter close to massive stars are privileged regions to target because of the prevailing high excitation conditions (strong UV radiation field and high gas density) which trigger dust evolution and also provide enough emission for detailed observations. The evolution of such regions and of more quiescent ones exposed to the standard interstellar radiation field \citep{Mathis1983} is due to the radiative heating, this is why they are called photon-dominated regions (PDRs). Bright PDRs such as the Orion Bar are active regions where the star formation process is best studied \citep{hollenbach1999}. Within the framework of an Early Release Science (ERS) observation program, the JWST space IR facility has observed gas and dust emissions towards the Orion Bar - the brightest PDR in the sky - at high angular resolution in imaging and spectroscopy \citep{berne2022ers,Habart2023jwst,peeters2023arXiv}, providing an ideal data set to study the evolution of dust and its nature. Using these data, we analyse the dust emission across the Orion Bar PDR and compare it to radiative transfer simulations in order to derive the effect of dust evolution.

The structure of the paper is as follows. In Sect.~\ref{sec:bar_pdr}, we provide an overview of previous studies of the Orion Bar. We recall the JWST spectroscopic and photometric observations in Sect.~\ref{sec:jwst_data}. In Sect.~\ref{sec:methods_tools}, we detail the THEMIS dust model and
the radiative transfer code used to simulate dust emission. 
 
We compare the modelled dust emission to observations and discuss the resulting constraints on dust properties in Sects.~\ref{sec:adjustments_nirspec} and \ref{sec:radiative_atomic_region}. Finally, in Sect.~\ref{sec:discussion},
we discuss our results and conclude in Sect.~\ref{sec:conclusion}.

\section{The Bar PDR}\label{sec:bar_pdr}

The Orion Bar, a strongly ultraviolet (UV) irradiated PDR, is an escarpment of the Orion molecular cloud (OMC), the closest site of ongoing massive star formation. The gas density in the ambient molecular cloud is estimated to be $n_\text{H}$~= 0.5--1.0~$\times$ 10$^5$~cm$^{-3}$ \citep{Tielens_1985b,Hoger95}.
The Bar is illuminated by the \mbox{O7-type} star \mbox{$\theta^1$ Ori C}, the brightest and most massive member of the Trapezium cluster at the heart of the Orion Nebula \citep[e.g.][]{Odell01}. The Trapezium cluster creates a blister \ion{H}{ii}, region that is gradually eroding into its parent cloud. A large cavity has been carved out of the molecular gas and the concave inner structure slopes to form the Orion Bar \citep{Wen1995, Odell01}. 

The edge of the Orion Bar PDR corresponds to the ionization front (IF), where the ionized gas (\ion{H}{ii}) recombines to neutral gas (\ion{H}{i}). Right after this edge, photons have energies below 13.6~eV and the incident far-UV (FUV) radiation field is $G_0$~= 2--7~$\times$ 10$^4$ times the Habing unit \citep[e.g.][]{peeters2023arXiv}. This atomic \ion{H}{i} layer extends over 10-20$\arcsec$ (or 0.02-0.04~pc) until dissociating FUV photons are sufficiently attenuated to let hydrogen become molecular \citep[e.g.][]{Habart2023jwst} across a dissociation H/H$_2$ front (DF). At this position the visual extinction to the IF is \mbox{$A_V$\,$\simeq$\,0.5-2~mag}. 
Beyond the DF, between \mbox{$A_V$~= 2} and 4~mag, the C$^+$/C/CO transition takes place  \citep[][]{Tauber95} and the PDR becomes molecular.

Recently the ALMA and JWST images at high angular resolution (\mbox{$\sim$0.1--1$\arcsec$}) have revealed small-scale ($\sim$0.004~pc) over-dense structures, sharp edges, and bright emission from embedded proplyds \citep[object \mbox{203-506};][]{Champion17}, possibly induced by the intense UV-radiation field \citep{Gorti02,Tremblin2012}. All these small-scale structures are superimposed over a rather homogeneous layer of atomic \ion{H}{i} gas that extends almost 0.01 pc from the PDR edge \citep{Habart2023jwst}.

Previous studies of dust evolution in PDRs relied on data with a limited angular resolution 
\citep[e.g.][]{compiegne2008,arab2012,mori2012,mori2014,pilleri2015,schirmer2022} that lead to significant broadening and confusion of the emitting structures. In particular, in their preliminary study of dust evolution in the Orion Bar based on Spitzer and Herschel data (resolution of a few arcseconds), \citet{schirmer2022} were only able to place upper and lower limits on the smallest size and abundance of nano-grains, respectively. We show below that a detailed modelling of high angular resolution JWST data ($\sim$ 100 times higher than that of former observations) allows us to obtain quantitative constraints on the size distribution of nano-grains. In addition, original constraints on the optical properties of these nano-grains are derived from the JWST MRS and NIRSpec spectroscopic data. 
This dust characterisation is important for understanding the impact of radiative feedback on the dynamics and chemistry of the ISM \citep[e.g.][]{schirmer2021}.

\section{Description of the JWST data used in this study}\label{sec:jwst_data} 

The photometric and spectroscopic data used in our study were obtained from the ERS project PDRs4All (\#1288) 'Radiative Feedback from Massive Stars as Traced by Multiband Imaging and Spectroscopic Mosaics'\footnote{PDRs4All \url{https://pdrs4all.org/}} \citep{berne2022ers}. 

\subsection{Observations}

The MIRI and NIRCam maps are presented in \citet{Habart2023jwst}, the MRS speçctra in \citet{chown2023}, and the \mbox{NIRSpec} spectra in \citet{peeters2023arXiv}.  An in-depth description of the reduction procedure for these data, including corrections for instrument signatures, background removal, calibration, rectification, and final combination, can be found in these three papers.
The fractional contributions of line, aromatic infrared bands (AIBs), and continuum emission to our NIRCam and MIRI images are examined in detail in the Science Enabling Product 4 article of Chown et al., in prep. 

\begin{figure*}
    \centering
    \includegraphics[width=\textwidth]{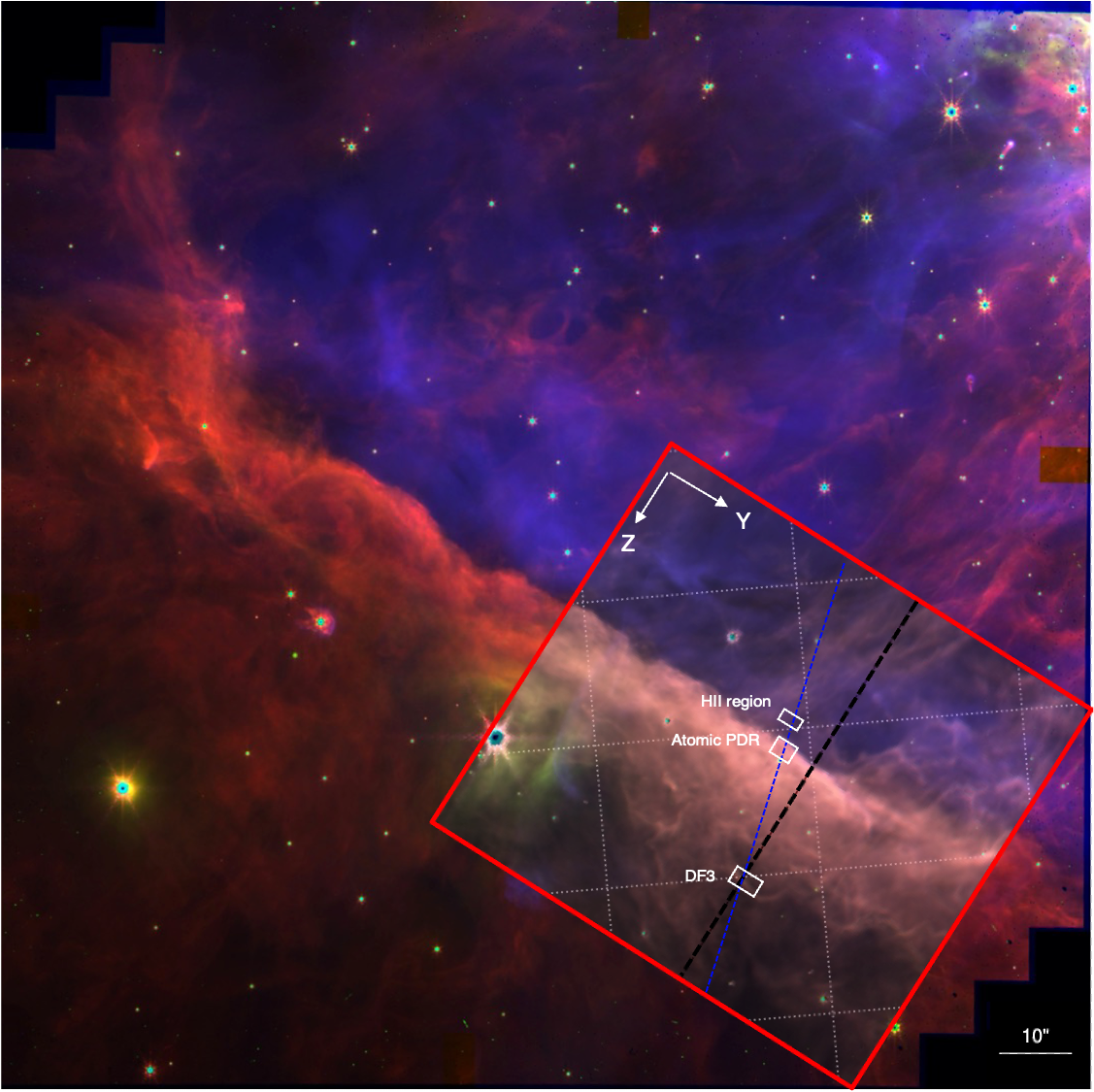}
    \caption{Inner region of the Orion Nebula as seen by the JWST's NIRCam instrument with north up and east left. This RGB composite image adapted from \citet{Habart2023jwst} is produced by combining three images in different filters: F335M (red), F470N (green), and F187N (blue) that trace emission from hydrocarbons (AIBs), dust and molecular gas (H$_2$ 0-0 S(9) line), and ionized gas (Pa $\alpha$ line), respectively. The red box delineates regions of particular interest zoomed in and shown in Figs.~\ref{fig:coupe1} and \ref{fig:coupe2}. The perpendicular (NIRSpec) cut is shown in black (blue) dashed line. The white boxes indicate the apertures used to extract the three template spectra from \cite{peeters2023arXiv}.  
  } \label{fig:Bar}
\end{figure*}

\subsubsection{Imaging data}

We use the F335M and F480M NIRCam filters. The F335M filter is dominated by the 3.3~$\mu$m AIB, with some contribution of the aliphatic bands at 3.4~$\mu$m
and the continuum. 
The F300M filter traces the emission continuum on the shorter wavelength side of the F335M filter, but it is affected by scattered light. 
We also use the F480M filter which is dominated by the continuum, the contribution of \HI\ lines, and H$_2$ line being negligible (Chown et al., in prep.).
We also use MIRI data in the F770W and F1130W filters, which include AIBs, and the F1500W and F2550W filters, predominantly capturing continuum emission.

\begin{figure*}[ht]
    \centering
    \includegraphics[width=0.8\textwidth]{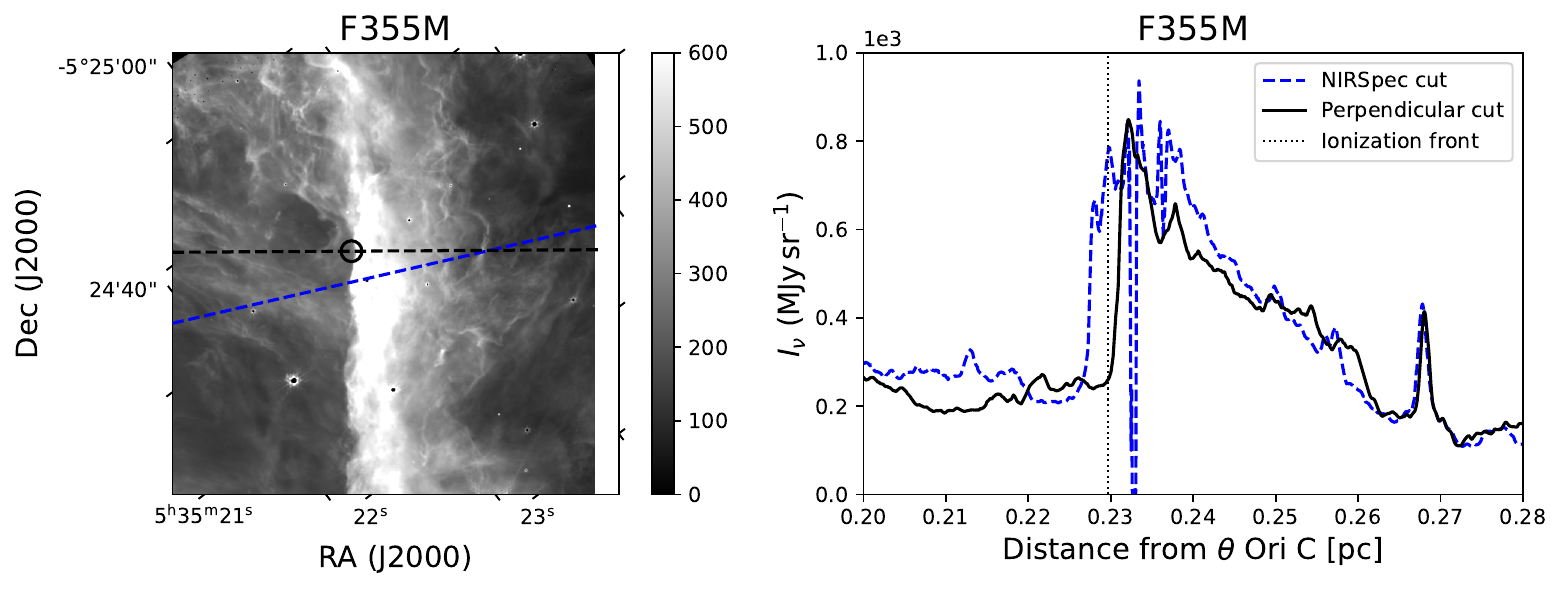}
    \includegraphics[width=0.8\textwidth]{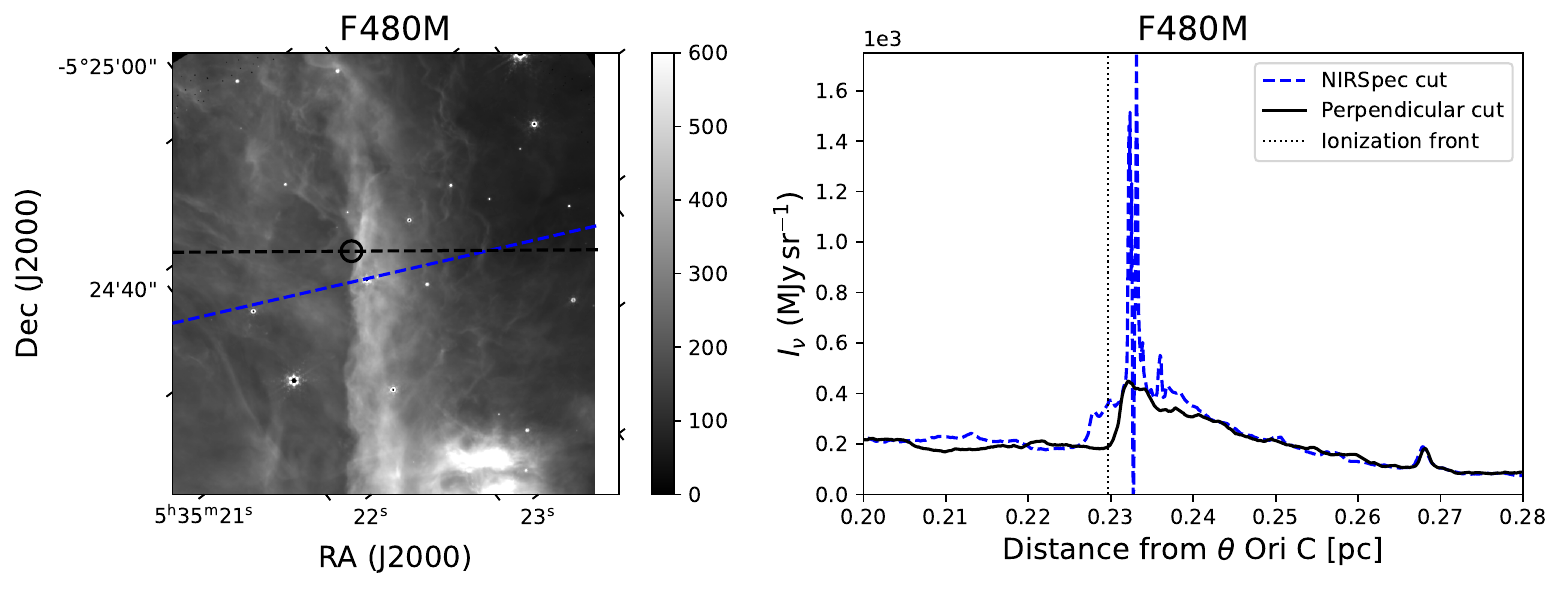}
     \includegraphics[width=0.8\textwidth]{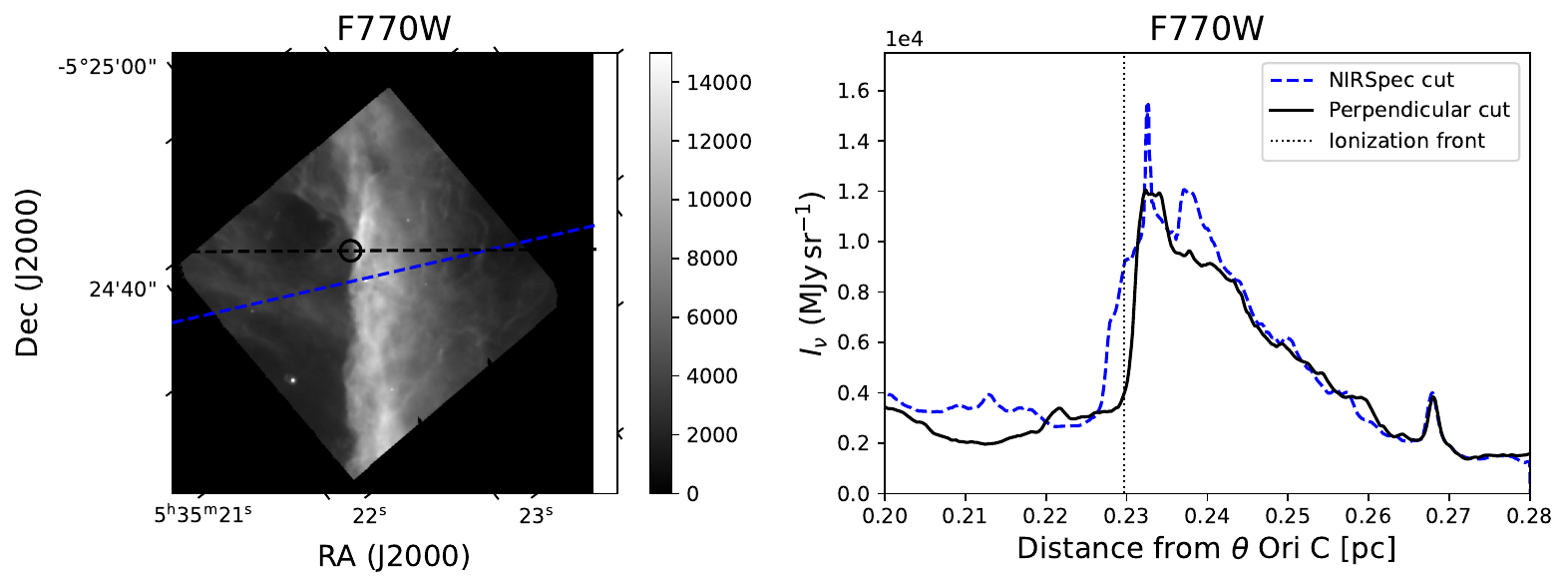}
    \caption{Spatial distribution of the NIRCam (F335M and F480M) and MIRI (F770W) filters tracing the AIBs at 3.35, 4.8, and 7.7~$\mu$m and continuum dust emission. {\bf Left:} Maps in the NIRCam (F335M and F480M) and MIRI (F770W) filters on a rotated Bar where the ionizing radiation is incident from the left. The blue inclined line shows the cut obtained from the NIRSpec field and the black line indicates the cut perpendicular to the Bar. A black circle shows the position of the IF for the perpendicular cut. {\bf Right:} Surface brightness profiles in the NIRCam and MIRI filters as a function of distance from the IF for perpendicular and NIRSpec cuts, with the average position of the IF marked by black vertical dash-dotted lines. Units are in MJy sr$^{-1}$. The observed very high pixel values are attributed to the NIRSpec cut crossing two proplyds, which affects the distribution of pixel intensities in this region.} \label{fig:coupe1}
\end{figure*}

\begin{figure*}[ht]
    \centering
    \includegraphics[width=0.8\textwidth]{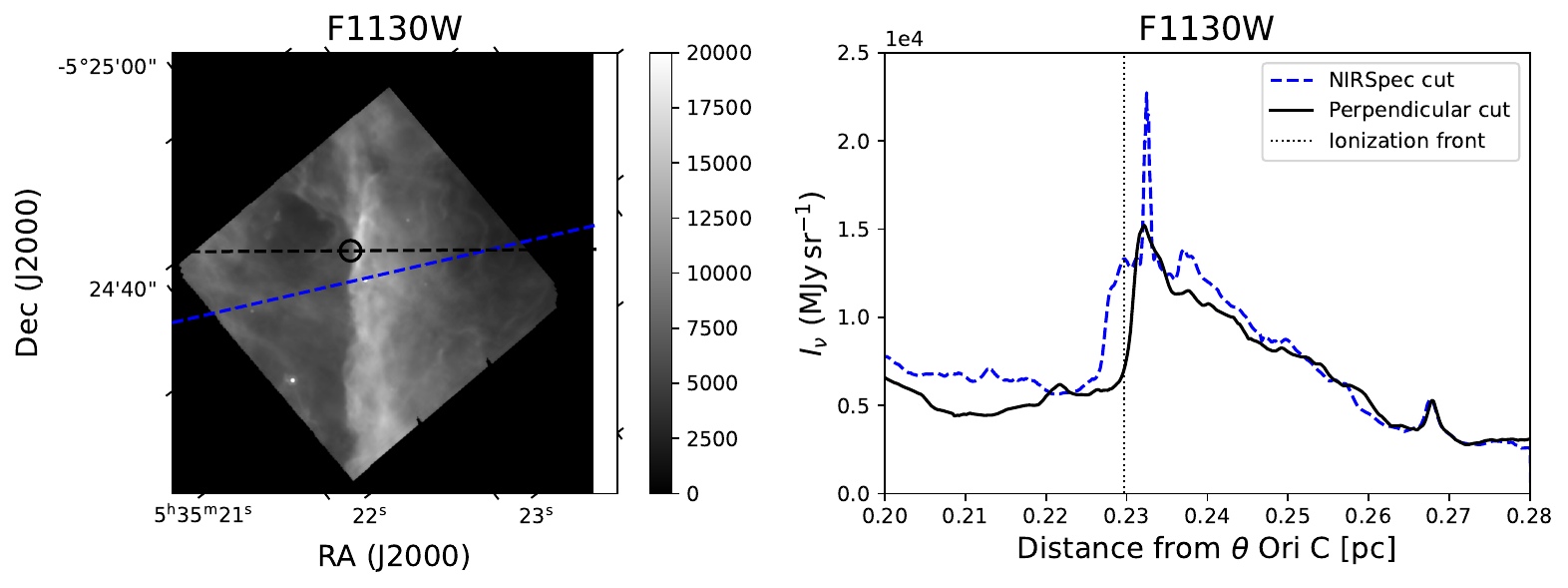}
    \includegraphics[width=0.8\textwidth]{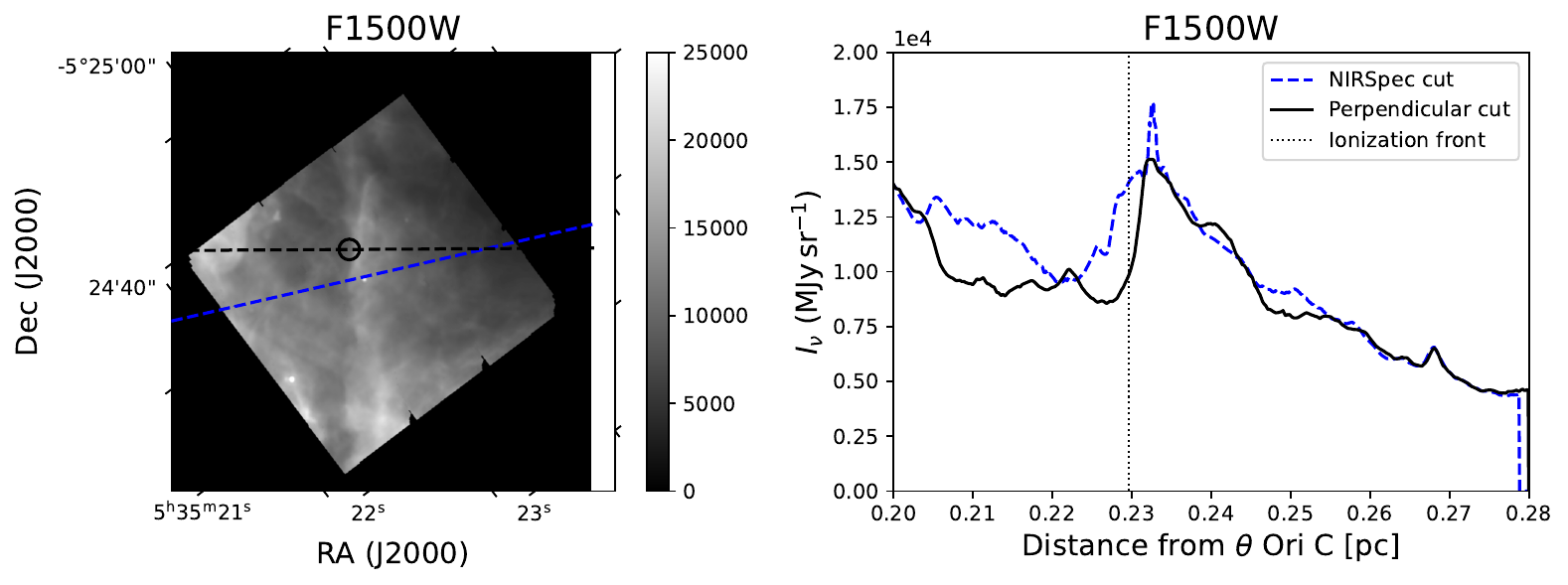}
     \includegraphics[width=0.8\textwidth]{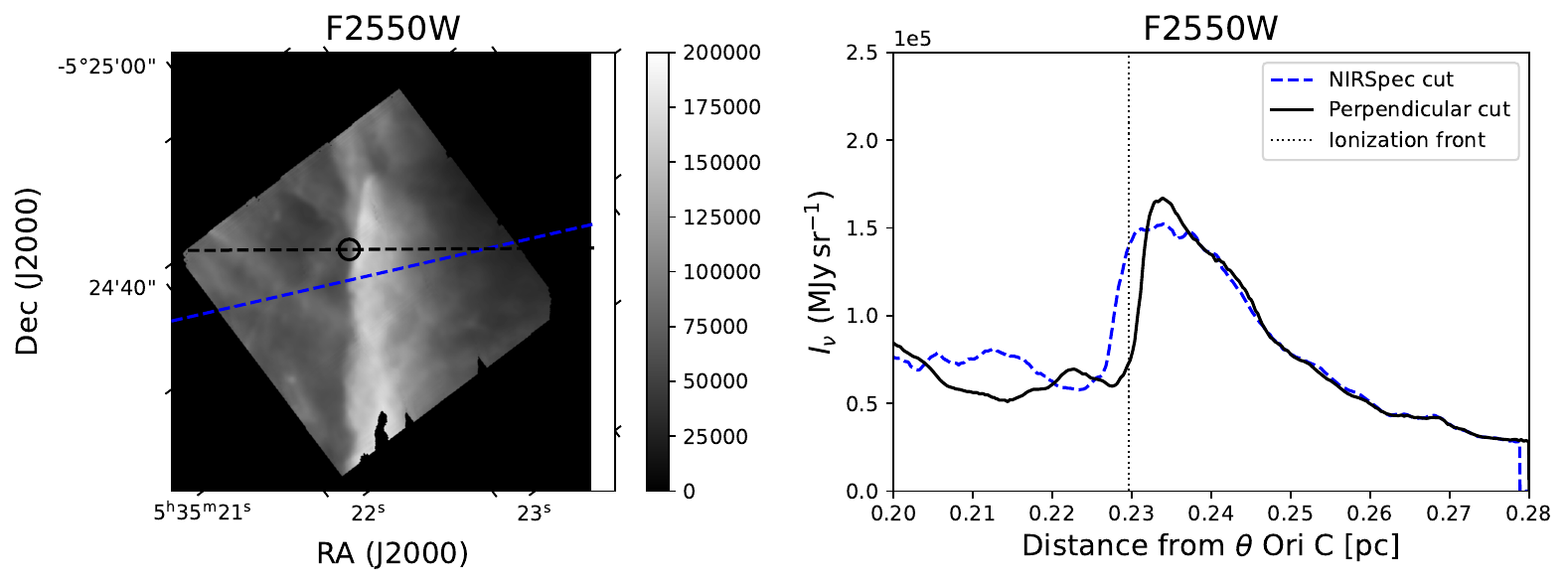}
    \caption{Same as Fig.~\ref{fig:coupe1}, but for MIRI filters F1130W, F1500W, and F2550W.} \label{fig:coupe2}
\end{figure*}

To showcase the data collected, Figs.~\ref{fig:coupe1} and \ref{fig:coupe2} present images with our selection of NIRCam (F335M and F480M) and MIRI filters (F770W, F1130W, F1500W, and F2550W). These filters trace the 3.3, 3.4, 7.7, and 11.3~$\mu$m bands alongside the continuum emission.
The F335M, F480M, F770W, and F1130W maps are dominated by the emission of very small carbonaceous grains, while larger grains contribute to the F1500W and F2550W filters. The contribution of the different grain sizes as a function of wavelength and radiation field strength is well illustrated in Figs.~8-10 of \cite{compiegne2011} and in Figs.~3 and 5 of \cite{schirmer2020}.

\begin{figure}
    \begin{center}
    \resizebox{\hsize}{!}{%
    \includegraphics{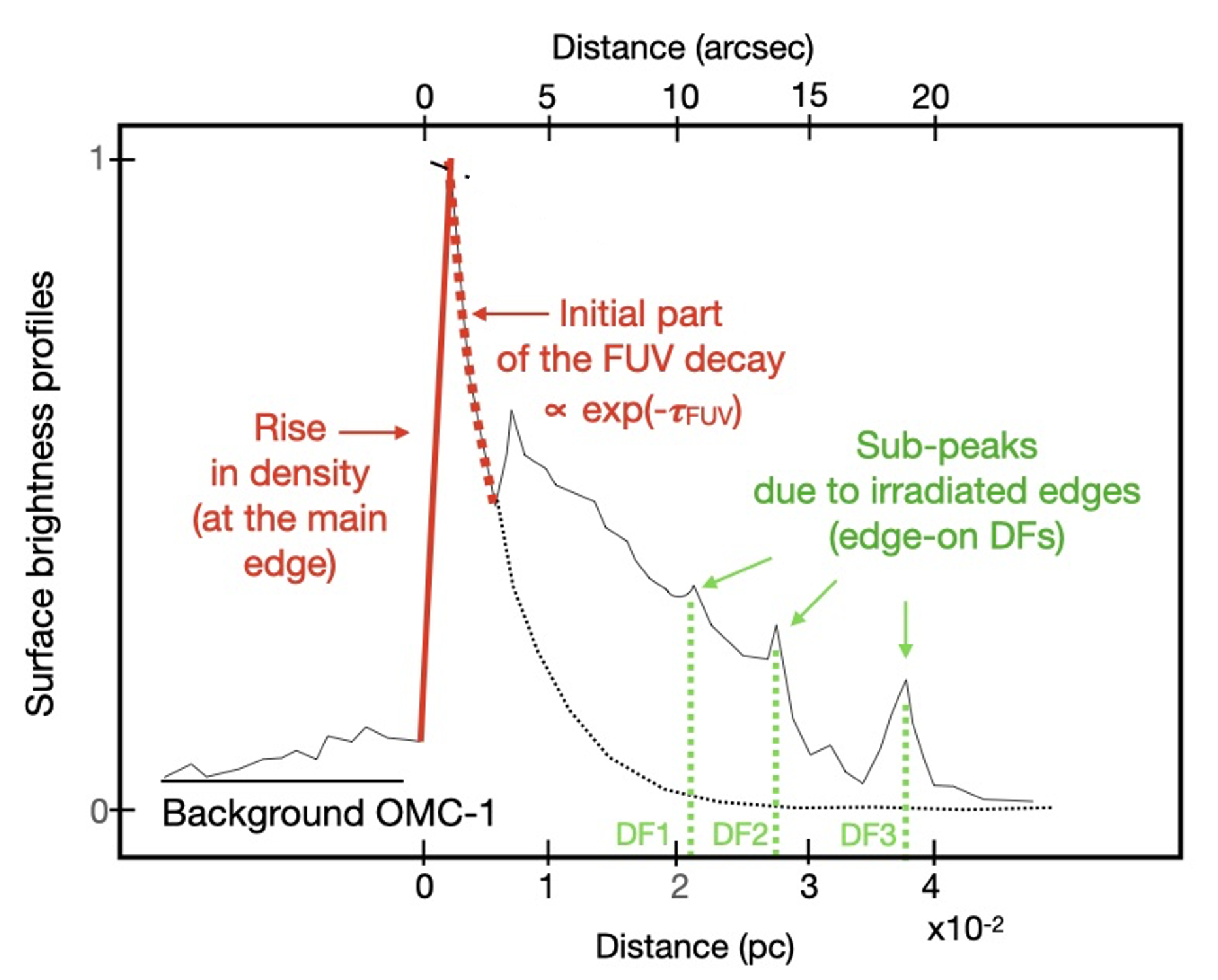}}
    \caption{Adapted from \cite{Habart2023jwst}. Schematic diagram illustrating the different components observed in the AIB surface brightness profile perpendicular to the Orion Bar. These components are shown for filter F335M (same as in Fig.~\ref{fig:coupe1}). In this study, we focus on modelling the first steep increase at the IF position, followed by a slower decrease indicated by the red line. The positions of DF1, DF2, and DF3 correspond to peaks in H$_2$ emission.
   } \label{fig:schema}
\end{center}
\end{figure}

The emission profiles perpendicular to the IF and the Bar are shown in Figs.~\ref{fig:coupe1} and \ref{fig:coupe2}, allowing us to probe dust emission variations across the PDR with increasing distance from the excitation source. 
These profiles provide insights into the density increase inside the PDR which occurs as the FUV radiation field decreases deeper into the Bar
\citep[see][and Fig.~\ref{fig:schema} adapted from their study]{Habart2023jwst}.

\subsubsection{Spectroscopic data}

As shown in their Fig.~1, \cite{peeters2023arXiv} extracted NIRSpec spectra from five strategically positioned apertures: in front of the ionization front (IF), at the peak of the 3.3~$\mu$m band, and at the \HI/\molh dissociation fronts (DF1, DF2, and DF3). These locations within the Orion Bar PDR are crucial as they represent the ionized, atomic, and warm molecular regions. \cite{chown2023} extracted MIRI template spectra using the same aperture positions as \cite{peeters2023arXiv}. By combining these MIRI spectra with the NIRSpec templates, we effectively captured the entirety of the AIB emission across the five regions of interest. In this paper, we specifically focus on three of the five combined NIRSpec and MRS template spectra representing key areas: the H II region, the atomic zone, and the DF3.

\section{Methods and tools}\label{sec:methods_tools}
In the following, we first explain the methodology employed for performing radiative transfer calculations. We then discuss the dust model we are using, which includes evolution from the diffuse interstellar medium to dense molecular clouds. Finally, we outline the assumptions made to represent the cloud geometry, the radiation field, and the free parameters, and an approach to constrain these parameters.

    \begin{figure}
        \centering
            \includegraphics[width=0.5\textwidth]{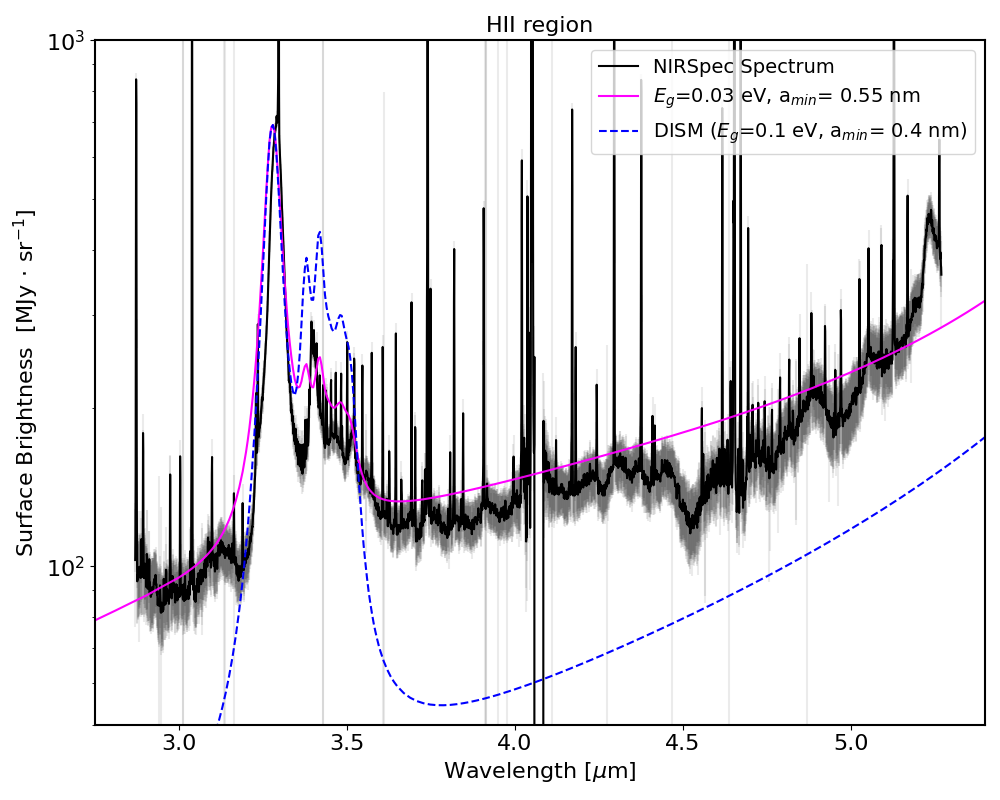}
            \includegraphics[width=0.5\textwidth]{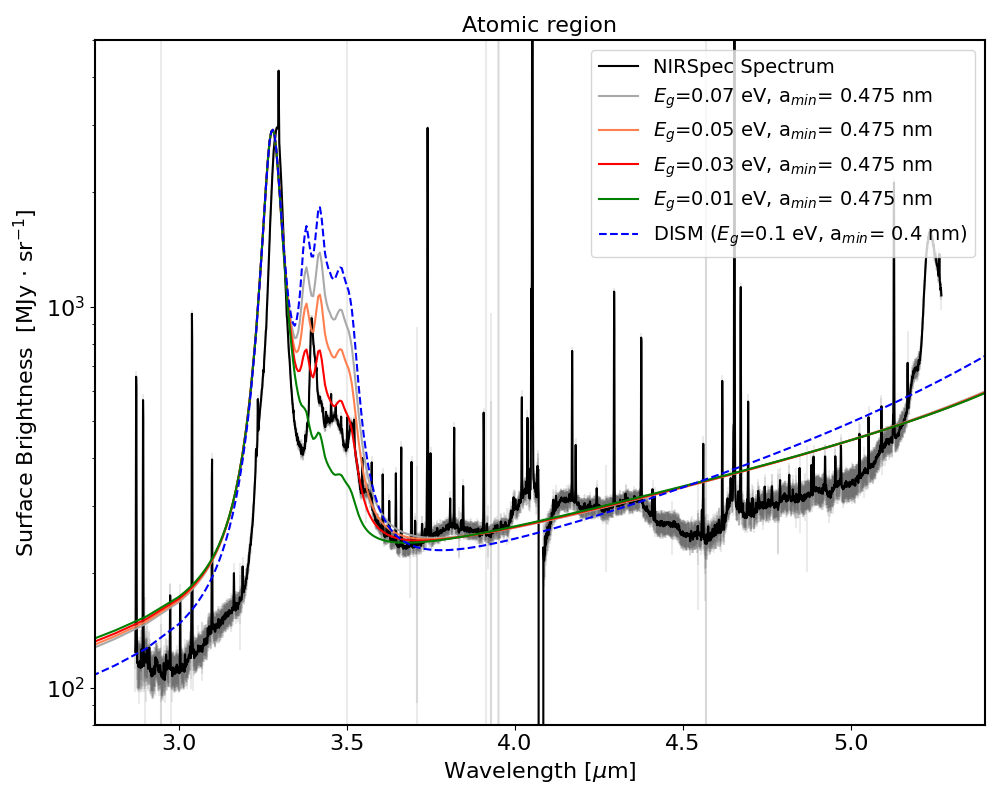}
            \includegraphics[width=0.5\textwidth]{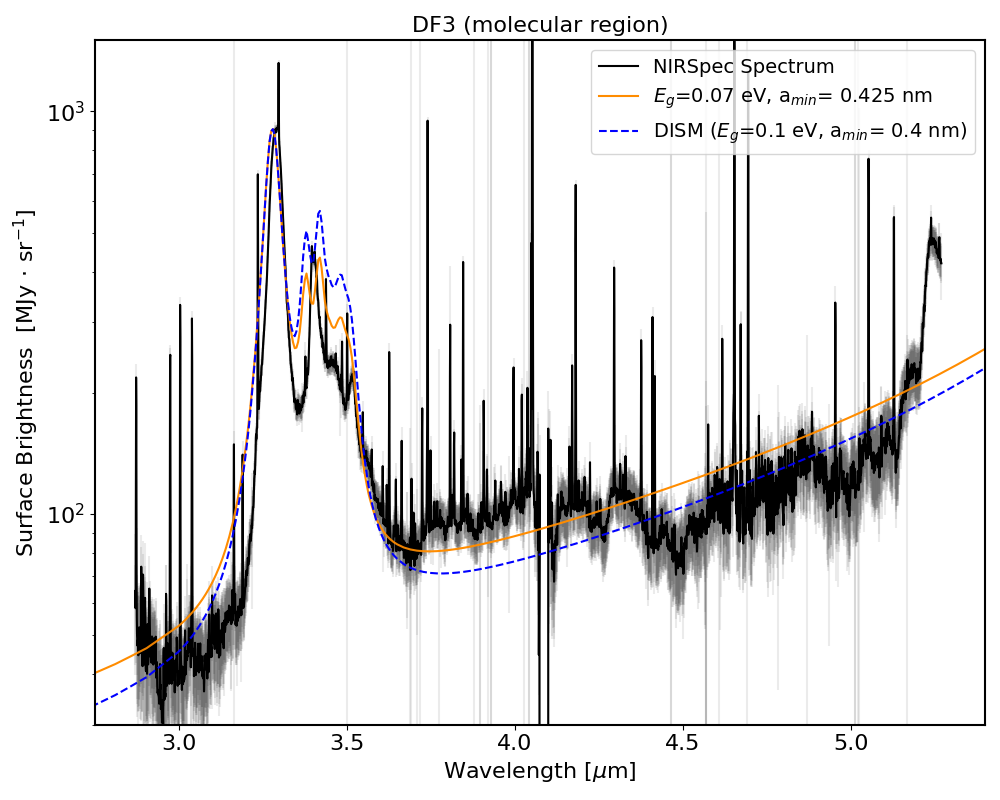}
    \caption{
    Comparison of observed surface brightness and DustEM models normalized to the 3.3~$\mu$m observed band (without radiative transfer) in the \ion{H}{ii} ({\bf upper}), atomic ({\bf middle}), and DF3 ({\bf lower}) regions. The black data points represent observations from NIRSpec, while the coloured lines show model predictions with varying band gap energies. In each figure, the DISM model is shown in blue dashed line and the best models for the \ion{H}{ii}, atomic, and DF3 regions are shown in magenta, red, orange respectively.}\label{fig:nirspec}
    \end{figure}
    
\subsection{Radiative transfer}\label{subsec:radiative_transfer}
We use the radiative transfer program \emph{Scattering with OpenCL} \footnote{SOC is available here: \href{https://github.com/mjuvela/SOC}{https://github.com/mjuvela/SOC}} \citep[SOC;][]{Juvela2019} which is based on the Monte-Carlo method, the simulation of large numbers of photon packages that represent the actual radiation field. This enables self-consistent modelling of the emission, radiation transport, and extinction throughout the model volume. This method relies on probability to determine the distance a photon travels before interacting with a grain, the type of interaction (scattering or absorption), and the direction of the photon in case of scattering. The calculations are performed with the coupling of the Monte Carlo radiative transfer code SOC and the dust emission and extinction code DustEM \footnote{DustEM is available at: \href{https://www.ias.u-psud.fr/DUSTEM}{https://www.ias.u-psud.fr/DUSTEM/}} \citep{compiegne2011}. Dust properties are defined by the absorption and scattering efficiencies $Q_\text{abs}$ and $Q_\text{sca}$, and the asymmetry parameter of the Henyey-Greenstein scattering phase function $g$, which are determined using the dust model defined in the next section. 

\subsection{Dust model description\label{subsec:themis}}
The Heterogeneous dust Evolution Model for Interstellar Solids\footnote{THEMIS is available here : \href{https://www.ias.u-psud.fr/themis/index.html}{https://www.ias.u-psud.fr/themis}} \citep[THEMIS;][]{jones2017} is built upon the foundations of the laboratory-measured properties of physically reasonable interstellar dust analogue materials. These include the family of hydrogenated amorphous carbon materials, from H-poor (a-C) to H-rich (a-C:H), collectively a-C(:H), and also amorphous olivine-type and pyroxene-type silicates with iron and iron sulphide nano-inclusions, a-Sil \citep{Jones2013,kohler2014}.

In the initial model for the diffuse ISM, small grains up to 20~nm in size are entirely made up of aromatic-rich, hydrogen-poor amorphous carbon (a-C), with a power-law size distribution and an exponential cut-off and optical properties that are coherent and continuous through the entire size range. These a-C grains are responsible of both AIBs, generally attributed to the astrophysical PAHs, and infrared continuum.

Larger grains have a core/mantle (CM) structure (a-C:H/a-C CM grains) with a-C mantles (depth $\sim$ 20 nm) enveloping a-C:H cores, and a-Sil/a-C grains with a-C mantles (depth $\sim$ 5--10~nm) \citep{Jones2013,kohler2014}. The two populations of larger grains have log-normal size distributions peaking between 100 and 200~nm. THEMIS takes into account the changes in dust properties through the effects of UV photo-processing, accretion, and coagulation, as the gas density evolves \citep{kohler15,jones2016c}. In the transition from the diffuse ISM to dense clouds or in environments with attenuated radiation or efficient rehydrogenation, the carbonaceous mantles can retain their hydrogen content, leading to the formation of grains with two mantles (core/mantle/mantle grains, CMM).  This formation can occur due to the coagulation of small aromatic-rich carbon grains onto larger grains and by the accretion of C and H atoms from the gas phase to form an a-C:H mantle. Further, within dense clouds, these CMM grains can coagulate into aggregates (AMM). Finally, the formation of ice mantles (I) on the aggregates (AMMI) can occur in the densest regions, shielded from energetic photons and gas molecules can freeze out onto the grain surfaces, possibly allowing chemistry to proceed \citep{kohler15}. 

Another key feature of THEMIS is the recognition of the inherent variability in the optical properties of hydrocarbon grains. As these or semi-conductor materials darken upon exposure to UV light and through thermal annealing, there is a decrease in the band gap or optical gap energy, $E_\text{g}$ \citep{iida1985,smith1984}. This property, and the corresponding effects on the optical properties, is crucial for understanding the evolutionary histories of hydrocarbon grains in the ISM \citep{duley1996,jones2009,jones2012a}. The band gap $E_\text{g}$ correlates linearly with the fractional atomic hydrogen content $X_\text{H}$ \citep[$\simeq$ $E_\text{g}$/4.3;][]{tamur1990}, and the band gap inversely depends upon the number of aromatic rings in the aromatic domains \citep{robertson1987}. A key element of the THEMIS framework is a self-consistent treatment of the evolution of the dust material properties (size distribution, chemical composition and structure) as they react to and adjust to the local radiation field intensity and hardness, and to the gas density and dynamics. For the carbonaceous nano-grains, all these processes, interactions will lead to changing of the C/H in the grains and also to changing of the dust structure for more or less aliphatic- or aromatic-rich. As detailed in \cite{jones2012a,jones2012b,jones2012c}, this is fully captured by the band gap in the optEC model.

In the forthcoming sections, we will test two variations of the THEMIS dust model. The dust populations in each case demonstrate specific size distributions and optical properties, indicative of the ISM's dynamic cosmic dust evolution.
They are:
\begin{itemize}
    \item THEMIS for the diffuse ISM (DISM), featuring two dust populations and three components consisting of a-C, a-C(:H)/a-C, and a-Sil/a-C. 
    \item The a-C + Mix(1:50) model, which substitutes THEMIS aggregates with spherical grains composed of 2/3 a-Sil and 1/3 a-C. These spherical grains exhibit a 50\% porosity level and a log-normal size distribution with a minimum size $a_\text{min,a-C}$ of 0.5~nm, a peak size $a_{0}$ (free parameter in the following) and a maximum size $a_\text{max}$ of 4.9~$\mu$m. This case extends the size distribution to larger grains, using the method proposed by \citet{Ysard2019} as shown in the left panel of Fig.~\ref{fig:size_dist}.
\end{itemize}

\subsection{Geometry and density profile}\label{subsec:geometry_density}
We take a plane-parallel geometry to represent an edge-on PDR, where all photons enter perpendicular to the PDR's edge. The cloud model is discretized onto a Cartesian grid consisting of $N_\text{X}$~$\times$ $N_\text{Y}$~$\times$ $N_\text{Z}$ cubes with each cube having a size of 0.001~pc (equivalent to 0.5$\arcsec$). This cell size is three times smaller than the smallest scales observed in the profiles (see Fig.~\ref{fig:coupe1}).
We set $N_\text{Z}$~= 100, which corresponds to the number of cubes along the PDR-star axis, and $N_\text{Y}$~= 10 (to avoid edge effects), corresponding to the number of cubes along the axis Y perpendicular to both the PDR-star axis Z and the line-of-sight axis X. The number of cubes in the PDR along the line-of-sight, $N_\text{X}$, depends on the value of the length of the Bar along the line-of-sight, $l_\text{PDR}$, given by the equation: $l_\text{PDR}$~= $N_\text{X}$~$\times$ cell size. This is a free parameter which will be constrained. 

We take the density profile used in previous gas and dust models of PDR edges \citep{habart05,arab2012,schirmer2020,schirmer2022}:

\begin{equation}
    \label{eq:density_profile}
    n_{\text{H}}(z)=\left\{ 
\begin{array}{l l}
  n_{0} \left(\frac{z}{z_{0}}\right)^{\gamma}  & \quad \text{if $z<z_{0}$}\\
  n_{0} & \quad \text{if $z>z_{0}$ ,}\\ \end{array} \right.
\end{equation}
where $z$ is the distance from the edge of the PDR. The power-law index $\gamma$ governs the steepness of the density front and is equal to 2.5. The maximum value of the density $n_{0}$ is reached at the depth $z_{0}$ and is assumed to remain constant for $z>z_{0}$. 

\begin{figure*}
    \centering
    \includegraphics[width=\textwidth]{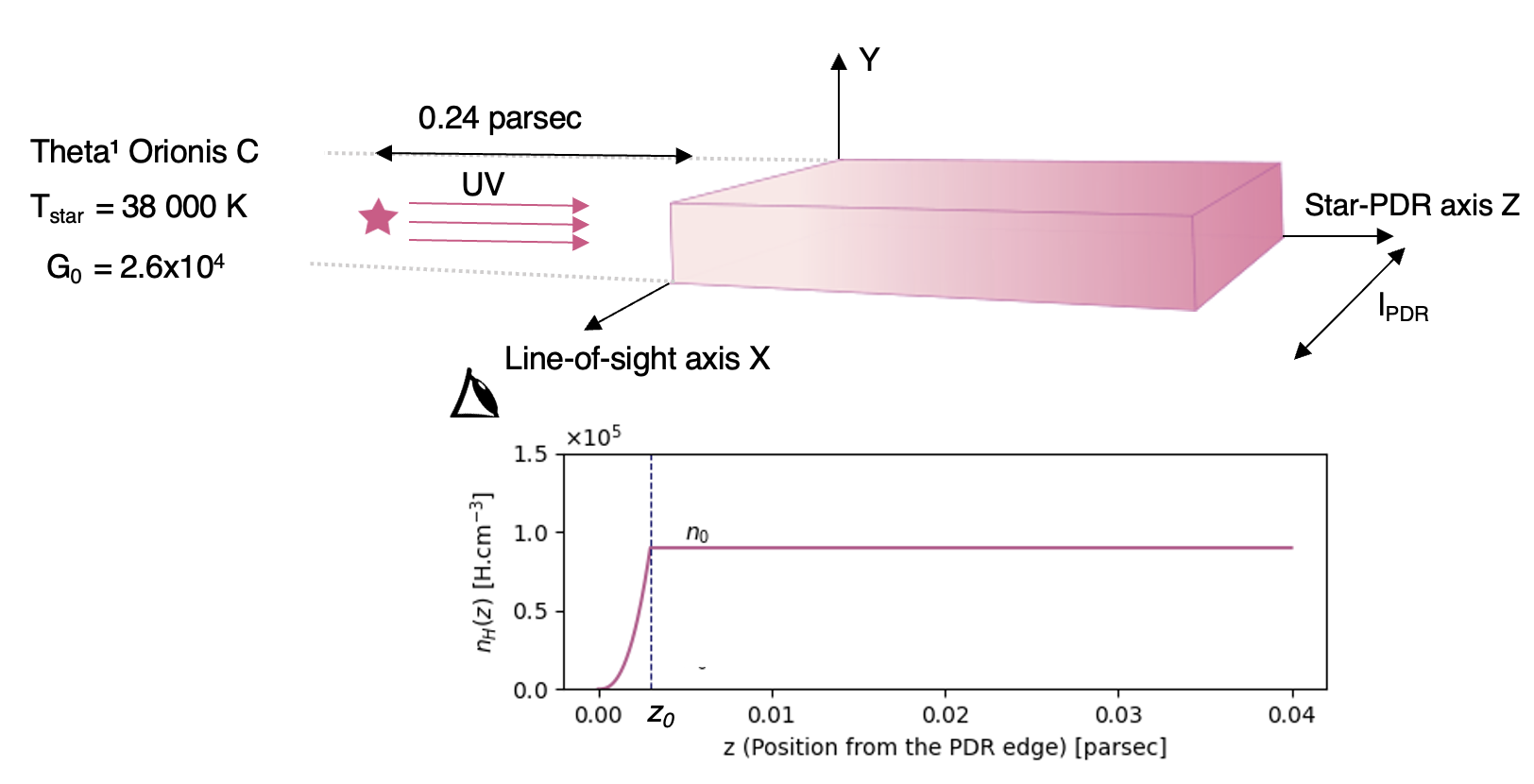}
    \caption{Schematic illustration of the PDR illuminated by a radiation field (from the left). {\bf Lower}: Assumed density profile across the Bar (pink line, see Sect.~\ref{subsec:geometry_density}). No constraints are given on density profile after $z$~= 0.01~pc (see Sect.~\ref{subsec:geometry_density}).}
    \label{fig:geometry}
\end{figure*}

\subsection{Radiation field}\label{subsec:radiation_field}

The PDR is illuminated by the \mbox{O7-type} star \mbox{$\theta^1$ Ori C} \citep{sota2011},
the most massive member of the Trapezium young stellar cluster, and its spectrum can be modelled by a $T$~= 38\,000~K blackbody.  We fix the incident radiation entering the PDR so that, at the
edge of the PDR, $G_{0}$~= 2.6~$\times$ 10$^{4}$  
in Habing units (in agreement with estimates from UV-pumped IR-fluorescent lines of \ion{O}{i} by \citealt{marconi1998} and \citealt{peeters2023arXiv} which indicate $G_0$~= 2.2--7.1~$\times$ 10$^4$).
The star was assumed to be at a distance of 0.24~pc from the PDR front, which corresponds to the projected distance between the PDR and the star (about 2$\arcmin$ north-west of the Bar, e.g. \citealt{Habart2023jwst}). This radiation field is calculated on a grid of 757 frequencies regularly sampled, except for frequencies close to bands where we use higher resolution in order to sufficiently resolve these features (see Fig.~\ref{fig:geometry} for a schematic illustration of the PDR).

\subsection{PSF convolution and band integration of our dust emission model}\label{subsec:psf}
In order to compare our dust emission models with observational data, it is crucial to perform a post-processing treatment of the model outputs, including  integration over specific photometric bands and convolution with the Point Spread Functions (PSFs) of the JWST.
The NIRCam and MIRI PSFs are taken from WebbPSF\footnote{\url{ https://github.com/spacetelescope/webbpsf}} \citep{Perrin20214}. Following the convolution, we integrate our model outputs over the different photometric bands using the photon-to-electron conversion efficiency. 
\subsection{Methodology and free parameters}\label{subsec:methodo_free}
Our methodology involves a grid-based analysis, where we explore a range of free parameters to construct a set of models using THEMIS and SOC. These models allow us to examine the influence of various parameters on the resulting dust emission profiles and compare them with photometric data from NIRCam and MIRI, as well as with spectra from NIRSpec and MRS.

The free parameters in our model\footnote{The model parameters are only free within the sense that a given set of coupled parameters (e.g. minimum size and band gap) determine the observable dust properties over a wide wavelength range (e.g. NIR to MIR or MIR to millimetre). One cannot therefore adopt arbitrary parameter values.} incorporate elements associated with both nano-grain and larger grain properties, as well as the density profile: 

\begin{enumerate}
    \item \textbf{Band gap energy ($E_\text{g}$) - hydrogenation state}: is completely determined by the electronic structure of a-C(:H) nano-grains and is primarily and strongly constrained by the spectral data. This parameter particularly affects the 3.3-to-3.4~$\mu$m band ratio and as such is the initial focus of our grid exploration using NIRSpec spectra.
    \item \textbf{Abundance ($M_\text{a-C}$/$M_\text{H}$)}: The a-C dust mass to gas ratio, representing the relative abundance of the smaller dust grains ($\sim$ sub 20 nm radius), which determines the global level of emission from the NIR to the MIR.
    \item \textbf{Minimum size ($a_\text{min,a-C}$)}: Denotes the smallest a-C nano-grain size in the distribution. This parameter impacts both the spectral shape and emission continuum slope.
    \item \textbf{Slope ($\alpha$)}: of the a-C power-law size distribution. Like $a_\text{min,a-C}$, this plays a significant role in determining the spectral shape and emission continuum slope.
    \item \textbf{Size $a_{\text{0,Mix(1:50)}}$}: Peak of the log-normal size distribution for the large grains (Mix 1:50). This mostly impacts the dust temperature and therefore where the SED peaks in the MIR to FIR wavelength range.
    \item \textbf{Distance ($z_0$)} and \textbf{density ($n_0$)} are the density profile parameters already defined in Eq.(\ref{eq:density_profile}). They influence the spatial emission profiles (steepness and width) for all of the NIRCam and MIRI filter fitting.  
    \item \textbf{Length of the PDR ($l_\text{PDR}$)} along the line of sight  does not affect the shape of the dust spectrum and is considered as a multiplying factor on the dust spectrum. The intensity increases linearly with $ l_\text{PDR}$ without altering the shape of the dust spectrum \citep[see][]{schirmer2020}.
\end{enumerate}

The influence of variations in the three parameters $M_\text{a-C}$/$M_\text{H}$, $a_\text{min,a-C}$, and $\alpha$ on both the dust size distribution and the associated spectra in the optically thin limit are shown in \citet{schirmer2020}. Our challenge here is 
to derive realistic self-consistent models compatible with the observations
in an eight-dimensional space defined by $E_\text{g}$, $M_\text{a-C}$/$M_\text{H}$, $a_\text{min,a-C}$, $\alpha$, $a_{\text{0,Mix(1:50)}}$, $z_0$, $n_0$, and $l_\text{PDR}$. To tackle this problem, we adopt a strategy that involves fixing or constraining certain parameters whenever possible. 
Among these, $E_\text{g}$ is particularly well constrained by NIR spectral data and thus forms the starting point of our grid exploration using NIRSpec spectra (see Sect.~\ref{sec:adjustments_nirspec}). By evaluating the spectral influence of various $E_\text{g}$ values, we are able to refine our exploration domain by narrowing down the parameter space. 
A similar approach is adopted for the parameters $a_\text{min,a-C}$ and $\alpha$, due to their significant impact on the spectra (band shape and continuum slope). 
With $E_\text{g}$, $a_\text{min,a-C}$ and $\alpha$ established, we proceed with a more targeted search, varying $z_0$ and $a_{0}$ within reasonable ranges (see Sect.~\ref{subsec:z0_a0}). 
This approach generates a grid of models, allowing us to reduce the range of parameter values and complexity of our exploration grid space (Sect.~\ref{subsec:final_grid}).
The resulting emission profiles are then compared with the observed photometric data, aiming to identify the parameter combinations compatible with the observed profiles. 

\section{Adjustments of NIRspec template spectra}\label{sec:adjustments_nirspec}

We begin by conducting simulations using the DustEM code without radiative transfer, in order to produce a grid of THEMIS SEDs for $G_{0}$~= 2.6~$\times$ 10$^{4}$. 
The grid covers variations in the minimum grain size ($a_\text{min,a-C}$) from 0.35 to 0.8~nm, in increments of 0.05~nm,  the power-law slope ($\alpha$) from $-$12 to $-$5 with a step size of 0.25, and the hydrogenation state ($E_\text{g}$) from 0.01 to 0.1~eV.  
The modelled SED grid is then compared to three NIRSpec spectral templates corresponding to different regions in front of and inside the Bar \citep{peeters2023arXiv}.
These three spectra, shown in Fig.~\ref{fig:nirspec},  correspond to (1) the ionized region in front of the Bar (top), (2) the atomic region of the Bar (middle), and (3) the warm molecular region (bottom), respectively.

In comparing the spectra, we note variations in the 3.3~$\mu$m band (associated with aromatics) and the 3.4~$\mu$m band (associated with aliphatics), as well as differences in the band-to-continuum ratio. We initially compare our model with the DISM THEMIS model (with a minimum grain size of 0.4~nm and a 0.1~eV band gap, corresponding to a hydrogenation of $X_\text{H}$~$\sim$ 0.02). Although the continuum fits relatively well for some regions, we find discrepancies in the 3.4~$\mu$m band and the overall continuum level for other regions. In the \ion{H}{ii} region (top panel of Fig.~\ref{fig:nirspec}), the continuum is underestimated by a factor of $\sim$2, while on the atomic Bar (middle panel of Fig.~\ref{fig:nirspec}), the continuum fits relatively well, but the 3.4~$\mu$m band is significantly overestimated (factor of $\sim$2). Regarding the molecular filament DF3 (bottom panel of Fig.~\ref{fig:nirspec}), the fit is reasonably good for the continuum and the bands\footnote{We note small differences in the model spectra compared to the observed spectra in the 3.4~$\mu$m region,  due to the fact that we have not fine-tuned the model. It is evident that some band widths and positions, as used to fit the DISM, are not quite consistent with this high excitation
PDR. Further theoretical and experimental work will be needed to resolve these small differences.}.

To address the discrepancies found for the \ion{H}{ii} and atomic regions, we vary the minimum grain size and the degree of hydrogenation through $E_\text{g}$, in order to adjust the spectra. Figure~\ref{fig:nirspec} illustrates the effect of each parameter on the modelled SED. As expected, increasing the minimum size of the smallest and therefore the hottest grains decreases the SED intensity at the shortest wavelengths, having an effect on both the bands and the continuum. The effect of decreasing the band gap is most visible in the 3.3/3.4 ratio as it corresponds to removing hydrogen atoms from the grain structure and thus to a decrease in the 3.4~$\mu$m aliphatic band (see also \citealt{jones2012c,Jones2013} for details). The 3.4~$\mu$m intensity is modelled by the integrated intensity and the best agreement is achieved with the following parameters:
\begin{itemize}
    \item For the \ion{H}{ii} region, $E_\text{g}$~= 0.03~eV and $a_\text{min}$~= 0.55~nm.
    \item For the atomic region, $E_\text{g}$~= 0.03~eV and $a_\text{min}$~= 0.475~nm.
    \item For the warm molecular region (DF3), $E_\text{g}$~= 0.07~eV and $a_\text{min}$~= 0.425~nm.
\end{itemize}

In order to adjust the ratios of the 3.3 and 3.4~$\mu$m bands and match the observed continuum level, it was necessary to decrease the band gap and increase the minimum size $a_\text{min}$ in the transition from denser shielded regions to more irradiated regions. This indicates that as they get closer to the star, the hydrogen content in the grain structure decreases, consistent with previous photo-processing studies \citep[see for instance][and references therein]{Jones2014}. Additionally, we found that in the irradiated regions, slightly larger grains are needed because the smaller ones are apparently more easily destroyed. 

It is important to highlight that in the \ion{H}{ii} region, the contribution of free-free emission is likely to be significant. This could account for the observed underestimation of the continuum level attributed to dust emission, as suggested by \citet{Haraguchi2012}. 
For the 3.3--3.4~$\mu$m bands and the continuum, we must also consider the significant influence of the background contribution from the Orion Molecular Core~1 (OMC-1) \citep[see][and references therein]{wen1995three,Odell01}. At the \ion{H}{ii} region, the observed emission encompasses both the gas and dust from the ionized region as well as the PDR face-on from OMC-1. However, our results indicate an evolution in both hydrogenation and grain size from DF3 to the \ion{H}{ii} region, suggesting that emission from the ionized medium is a contributing factor. This observation underlines the complexity of the \ion{H}{ii} region's emission profile, where multiple sources and processes are at play.

Now and hereafter in Sect.~\ref{sec:radiative_atomic_region}, we will focus on the PDR edge, that is the atomic region. To best determine the range of possible $E_\text{g}$ values in this region model results for different band gap energies ($E_\text{g}$ = 0.01, 0.03, 0.05, and 0.07~eV) are compared to the observed data in the middle panel of Fig.~\ref{fig:nirspec}. This comparison highlights that the $E_\text{g}$~= 0.03~eV model best reproduces the observations in the \Tb region, closely following the trend of the data points and accurately reproducing the ratio between the 3.3 and 3.4~$\mu$m bands. In contrast, the models with $E_\text{g}$~$\geq$ 0.05~eV overestimate the intensity in the 3.4~$\mu$m band, while the model with $E_\text{g}$~= 0.01~eV underestimates the intensity in the same band. Thus, we assume a fixed value of $E_\text{g}$~= 0.03~eV for the nano-grain band gap. Moreover, based on the comparison between our models and the NIRSpec spectra in the atomic region, we found that the parameter space for $a_\text{min,a-C}$ and $\alpha$ to be explored to fit the NIRCam and MIRI data is the following: $0.4$~$\leqslant$ $a_\text{min,a-C}$~$\leqslant$ 0.6~nm, $-6$~$\leqslant$ $\alpha$~$\leqslant$ $-4$. 

\begin{figure}
\centering
\includegraphics[width=0.5\textwidth]{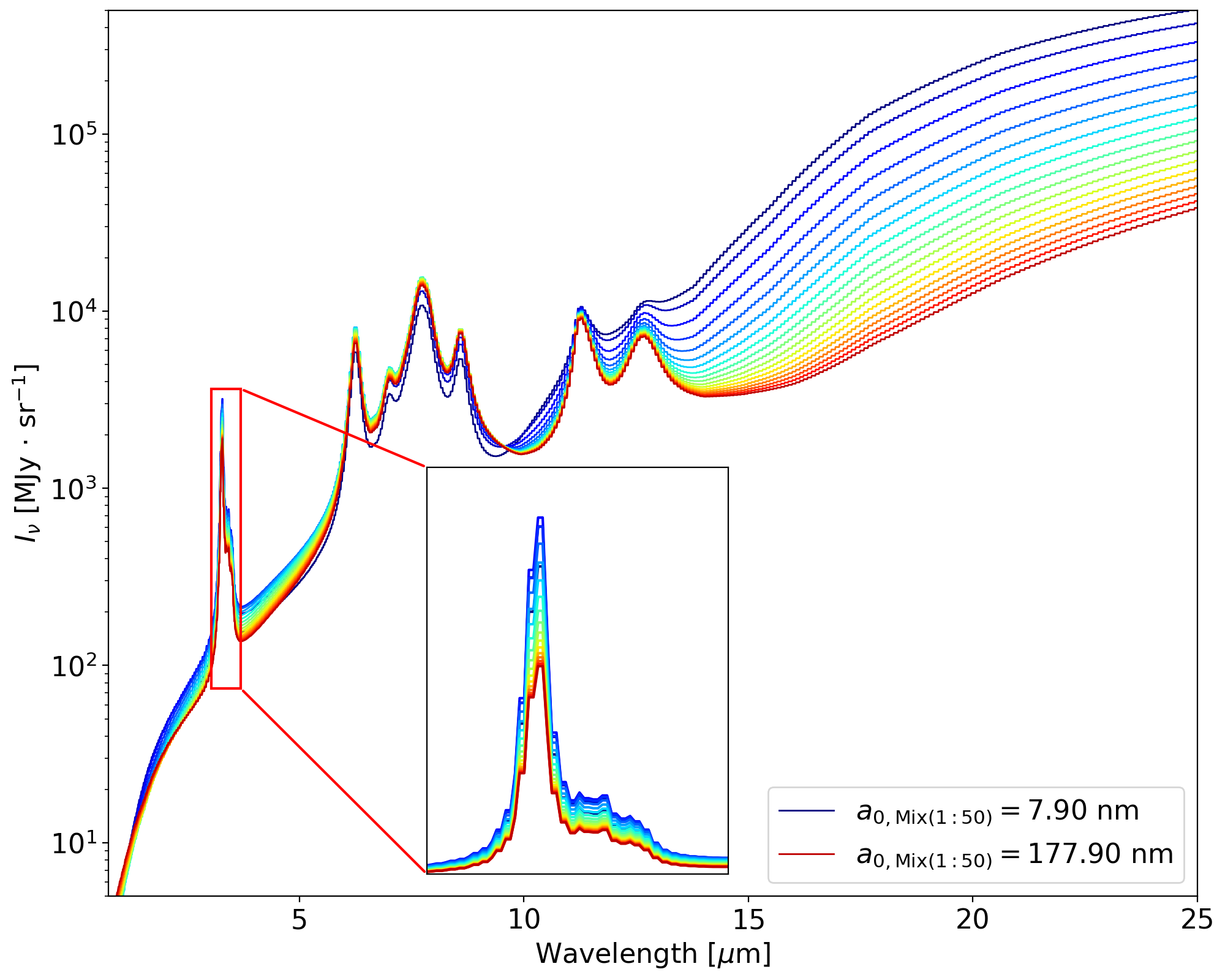}
\caption{Dust spectra for different values of $a_\text{0,Mix(1:50)}$ from 7.90 to 177.90~nm. The spectra are computed with radiative transfer using the SOC program and convoluted at the resolution of NIRSpec and MRS.
}
\label{fig:a0mix150}
\end{figure}

\begin{figure*}
    \centering
    \includegraphics[width=0.95\textwidth]{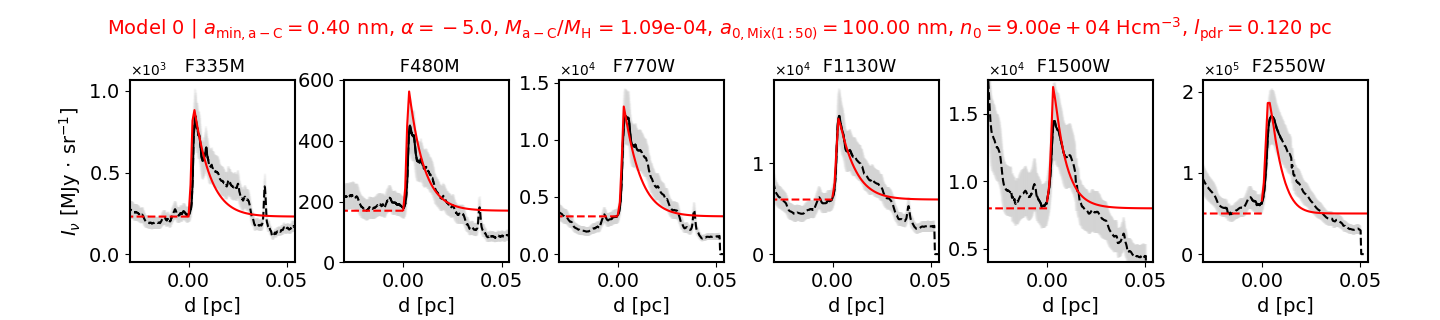}
    \includegraphics[width=0.95\textwidth]{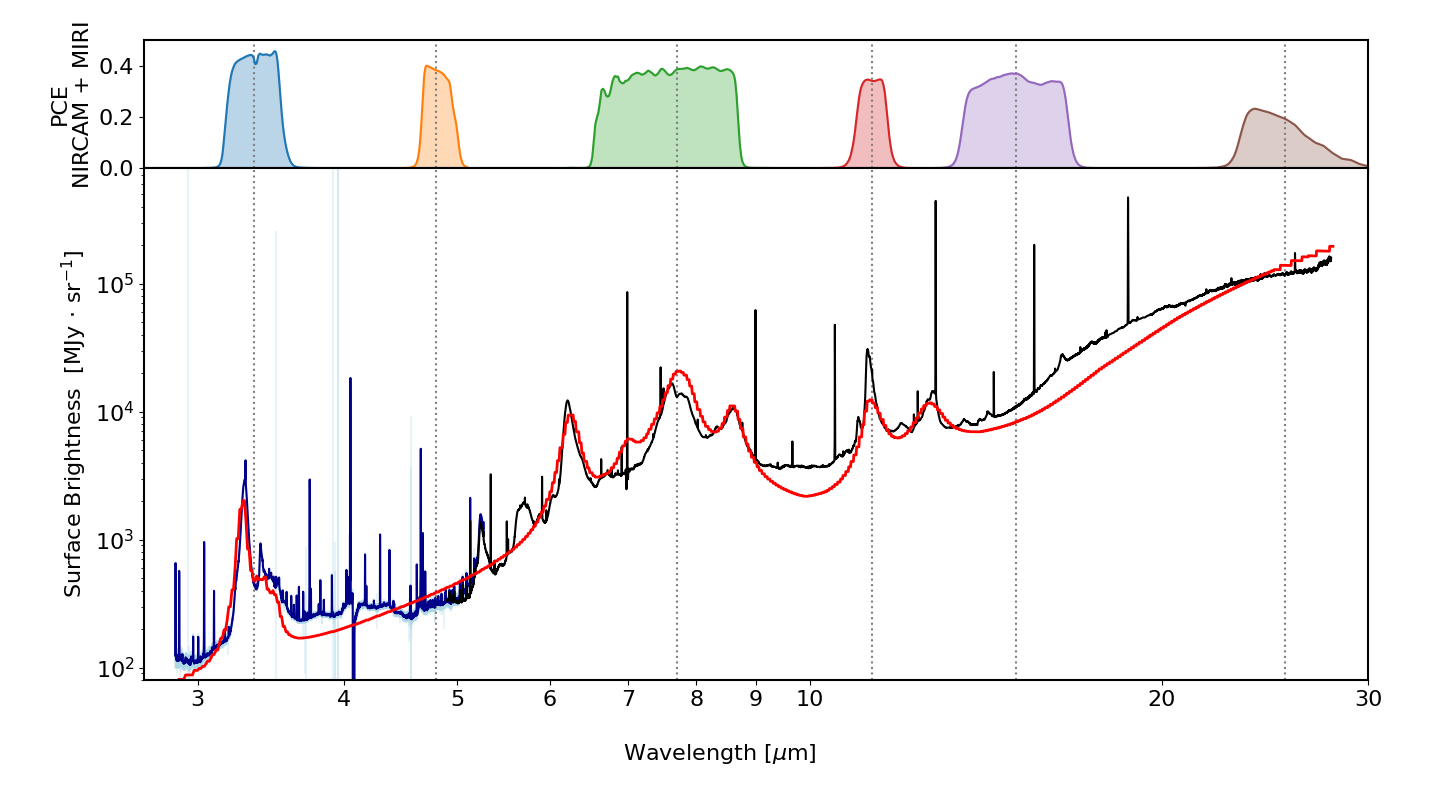}
    \caption{Comparison of the observed (JWST) and modelled dust emission in the atomic PDR region of the Orion Bar. \textbf{Top:} Comparison of the observed and modelled dust emission profiles in the \Tb region of the Orion Bar, using the best set of dust parameters in six photometric bands (3.3 and 4.8~$\mu$m; NIRCam filters, and 7.7, 11.3, 15, and 25.5~$\mu$m; MIRI filters). The observed dust emission is shown in black and the model fit in red, with uncertainties of 20$\%$ displayed in grey. The observed emission that we model (indicated by the first steep increase at the IF position, followed by a slower decrease) is shown in solid black line, while the other components of the observed emission are represented by dashed lines as explained in Fig.~\ref{fig:schema}. The cut is perpendicular across the Orion bar, as shown in Fig.~\ref{fig:Bar}.  \textbf{Middle:} Wavelength coverage and photon-to-electron conversion efficiency (PCE) of the NIRCam and MIRI imager filters used in our study. Throughput refers to photon-to-electron conversion efficiency. The central wavelengths of the filters are indicated with dashed vertical lines. The name of each filter is indicated in the top figure. The values of central wavelengths and bandwidths are presented in \citet{Habart2023jwst}. \textbf{Bottom:} Comparison of the dust emission spectral energy distribution (SED) calculated using our radiative transfer model at the AIB peak emission at a distance of 0.003~pc (represented by the red line), with template spectra for the \Tb in the Orion Bar from \citet{peeters2023arXiv} (shown in dark blue for NIRSpec and black for MRS) in the wavelength range of 2.75 to 28~$\mu$m.  }
    \label{fig:bestmodels0}
\end{figure*}

\section{Radiative transfer modelling in the atomic region}\label{sec:radiative_atomic_region}

As part of our comprehensive approach, we have constructed a model grid for NIRCam and MIRI photometric data in order to model the brightness profiles observed at the edge of the PDR, in particular, the rise in emission just after the IF and the decay due to extinction\footnote{The extended emission and secondary peaks towards the molecular region at the DFs are not modelled here. The extended emission is in part due to material along the line of sight located in front of the PDR, in the foreground face-on surface layer \citep[as seen in geometry on Fig.~5 of][]{Habart2023jwst}. This emission may originate in the compressed region that is still illuminated directly by the ionizing stars.}. This zone is important because it is where the density of the atomic gas increases sharply and the extinction (in the UV illuminating direction) perpendicular to the Bar begins. In this atomic zone, the dust significantly contributes\footnote{Based on the PDR Meudon code, the photoelectric effect along with the heating by H$_2$ cascades (vibrational de-excitation) will dominate the heating in the pure atomic region of the Orion Bar.} (at least 50\% for a density $n_0$~$\leq$ 1~$\times$ 10$^5$) to the gas heating via the photoelectric effect and it is crucial to provide constraints on its properties. The PDR edge was unresolved in the IR until now. Thanks to the JWST, we are now able to spatially resolve this zone, and we can put constraints on the dust properties.

\subsection{Constraining $z_0$ and $a_{\text{0,Mix(1:50)}}$ \label{subsec:z0_a0}}

One of the key parameters we aim to constrain in our analysis is the depth threshold ($z_0$) where the density profile reaches its maximum $n_0$. The determination of this parameter is largely independent of the dust properties, and is primarily reliant on the width of the dust-modelled emission. 
A value of $z_0$~= 0.003~$\pm$ 0.0005~pc is necessary to fit the brightness profiles. We note that $z_0$ is much smaller than the value of 0.025~pc deduced from the modelling of Spitzer and Herschel observations due to their lower spatial resolution \citep[][]{schirmer2022}. 

Another important parameter is the size of the pseudo-aggregates, $a_{\text{0,Mix(1:50)}}$. The variation of $a_{\text{0,Mix(1:50)}}$ of the log-normal size distribution of the large grains has a significant effect on the dust SED at all wavelengths.  This is because the dust size distribution 
affects the opacity from the visible to the submillimetre, which in turn affects the amount of radiation that is absorbed and re-emitted by dust. Figure~\ref{fig:a0mix150} shows the SED variations with $a_{\text{0,Mix(1:50)}}$. Increasing $a_{\text{0,Mix(1:50)}}$ broadens the log-normal grain size distribution. This results in a higher abundance of large grains, which are more efficient at absorbing long-wavelength radiation (IR) and a smaller population of small dust grains, which are more efficient at absorbing short-wavelength radiation (UV).
As a result, increasing the value of $a_{\text{0,Mix(1:50)}}$ will tend to shift the peak of the SED to longer wavelengths, as the dust will be cooler. 
Regarding our imaging data, this parameter has a significant influence on the ratio between the emission in the 15 and 25.5~$\mu$m bands.
Through an exploration of our preliminary grid, we concluded that a mean grain size ($a_{\text{0,Mix(1:50)}}$) of 100~nm most accurately produces the brightness profiles observed at 15 and 25.5~$\mu$m.  

\subsection{Remaining free parameters}\label{subsec:final_grid}

With $E_\text{g}$, $a_\text{0,Mix(1:50)}$, $z_0$ now established, the remaining parameters to determine are $a_\text{min,a-C}$, $\alpha$, $M_\text{a-C}$/$M_\text{H}$, $n_0$, and $l_\text{PDR}$. To explore these five parameters, we examined a more extensive model grid, spanning the following values:

\begin{enumerate}
    \item $M_\text{a-C}$/$M_\text{H}$ varies from 0.0025~$\times$ 10$^{-2}$ to 0.17~$\times$ 10$^{-2}$ on a 10-point linear grid. 
    \item $a_\text{min,a-C}$ varies from 0.4 to 0.6~nm on a 10-point linear grid. 
    \item $\alpha$ varies from $-$4.0 to $-$6.0 in steps of 0.2.
    \item $n_0$ varies from 4~$\times$ 10$^{4}$ to 1.5~$\times$ 10$^{5}$~cm$^{-3}$ in steps of 10$^{4}$.
     \item $l_\text{PDR}$ varies from 0.08 to 0.14~pc in steps of 0.01~pc.
\end{enumerate}

The  maximum value for $M_\text{a-C}$/$M_\text{H}$ corresponds to the value found in DISM.
The range of $a_\text{min,a-C}$ and $\alpha$ are discussed in Sect.~\ref{sec:adjustments_nirspec}.
Concerning $n_0$ and $l_\text{PDR}$, the range is taken to include the estimates from \cite{Habart2023jwst} and \cite{peeters2023arXiv}. In any case,  we have checked that for the five parameters, all values outside these ranges do not allow us to reproduce the data.
 
We are actually using an early stage of data reduction, which may have some calibration errors. 
In addition, the 3D model of the illuminated edge of the Bar is extremely simple, with a plane-parallel geometry observed edge-on, and illuminated by a star in the plane of the sky. Therefore a $\chi^{2}$ analysis to derive the 'best' model is premature. Instead, we opted for a grid search across the five parameters to be constrained ($M_\text{a-C}$/$M_\text{H}$, $a_\text{min,a-C}$, $\alpha$, $n_0$, $l_\text{PDR}$) in order to derive a set of parameters physically coherent and which provides a satisfactory agreement between a self-consistent model and the data (brightness profiles and spectra). 

We conducted multiple grid tests, varying the five parameters. Initially focusing on the DISM population, it was not possible to adjust the ratio between the 15~$\mu$m and 25.5~$\mu$m bands. Consequently, we explored a second dust population: the a-C + Mix(1:50) model (see Sect.~\ref{subsec:themis}).  The model best parameters were selected in order to minimize the quadratic difference between observed and modelled peak intensities across all six bands. Numerous models were tested (combinations of the 5 free parameters), and the only model that simultaneously reproduced all bands within the uncertainty range is the one presented in Sect.~\ref{subsec:adjust_photometry}. Other models, like the one in Fig~ \ref{fig:size_dist}, either overestimated or underestimated at least one band. 
Since our analysis encompassed all parameters of the dust model, we therefore believe that the simple geometry of the model is its primary limitation.

\subsection{Comparison with NIRCam and MIRI brightness profiles}
\label{subsec:adjust_photometry}
We took into account an offset for the observed emission which is crucial for the adjustment process.
Specifically, the value of the offset is taken in the \ion{H}{ii} region, just in front of the illuminated edge for each profile (shown as a dashed red line, top panel of Fig.~\ref{fig:bestmodels0}). This offset may have several physical origins (OMC-1 seen face-on or \ion{H}{ii} material, see Sec.~\ref{sec:adjustments_nirspec}) and its modelling is beyond the scope of this paper.  Additionally, we incorporated 20$\%$ error bars to account for the calibration accuracy. 

The model from our grid which provides the best agreement with the six brightness profiles is presented in Fig.~\ref{fig:bestmodels0}. The a-C abundance $M_\text{a-C}$/$M_\text{H}$ (0.0109~$\times$ 10$^{-2}$) is found to be 15 times less than in the diffuse ISM (0.17~$\times$ 10$^{-2}$). The minimum grain size $a_\text{min,a-C}$~= 0.4~nm and the slope $\alpha$~= $-$5 are consistent with the values found in the diffuse ISM \citep{Jones2013}. 
The gas density $n_0$ is estimated to be about 9~$\times$ 10$^{4}$~H~cm$^{-3}$ at the nano-grain emission peak.
This is in agreement with estimates made by \cite{Habart2023jwst} and previous estimates from atomic gas FIR lines (e.g. [\ion{C}{ii}] and [\ion{O}{i}] fine-structure lines, \citealt{bernard-salas2012}) and
estimates from the Raman-scattered wings
of hydrogen H$\alpha$ lines in the Bar
\citep{Henney2021}. Interestingly, this density is also close to the one derived at the DFs from NIRSpec data  \citep{peeters2023arXiv}. The value of $l_\text{PDR}$~= 0.12~pc we find is consistent with the value reported by \cite{peeters2023arXiv}. 
 
Notably, for the first time, we are able to simultaneously fit the profiles in all photometric bands (3.35 and 4.8~$\mu$m NIRCam filters and 7.7, 11.3, 15, and 25.5~$\mu$m MIRI filters) within the calibration accuracy. This achievement is significant because the bands at 3.35, 4.8, 7.7, and 11.3~$\mu$m cover both aromatic and aliphatic features and also the continuum, whereas the 15 and 25.5~$\mu$m bands only cover the dust continuum. The successful reproduction of the band-to-continuum ratio and band shapes in the Orion Bar suggests that the THEMIS model also works well in highly irradiated regions, despite the fact that it was constructed to explain DISM dust. 

\subsection{Comparison with NIRSpec and MRS spectra}\label{subsec:bestmodel_nirspec}

Using the derived parameters, the bottom panel of Fig.~\ref{fig:bestmodels0} shows the computed spectrum at the distance of 0.003~pc from the PDR edge which corresponds to the peak emission in all 
photometric bands. In Fig.~\ref{fig:bestmodels0}, we overplot the NIRSpec and MIRI spectra obtained within the atomic region of the Bar. Although the modelled and observed spectra are not obtained at exactly the same position on the sky (there is a slight offset of a few arcseconds, see Fig.~\ref{fig:coupe1}), both trace material in the atomic region of the Bar, allowing for a direct comparison.

The comparison between the modelled and observed spectra shows a good agreement both in the NIR and mid-IR part of the SED, covering the range from 3 to 25~$\mu$m. This global dust modelling approach is promising and highlights the effectiveness of our approach. However, we also notice some differences between the model predictions and the observed data:

1. The lack of emission features around 5--6~$\mu$m. 
\citet{jones2012a,jones2012b,jones2012c} assigned bands to the various C-C and C-H bonds according to the data available at that time and only bands with well-determined origins can be included within THEMIS. This means that an update of the positions and strengths of the various bands will need to be made in the light of the JWST observations once more spectra of regions with various $G_0/n_\text{H}$ will become available. This update will have to be based on these new observational constraints and, particularly, on new laboratory data and band assignments that build upon the available laboratory data \citep{dartois2004,pino2008,carpentier2012}.
 
2. The model tends to overestimate the continuum around 7~$\mu$m, between the 6.2 and 7.7~$\mu$m features. This is again related to the need to refine the band assignments based upon future laboratory data.

3. Although we can accurately reproduce the features at 3.3--3.4~$\mu$m in our model, we encounter challenges in adjusting the continuum around 3.5~$\mu$m. To address this, we consider the possibility of slightly increasing the value of $a_\text{min,a-C}$, which leads to an improved fit between 3 and 4~$\mu$m as shown in Fig.~\ref{fig:bestmodels2}. However, such an adjustment results in a slightly less accurate fitting of the F480M profile.  

4. 
The observations present a plateau above our model between 9 and 11~$\mu$m. This could be related to the presence of nano-silicates \citep[e.g.][]{Hensley2021,Chastenet2021,Zeegers2023} which are not included in THEMIS. For such small grains, stochastic heating would be efficient \citep{Guiu2022}, possibly leading to observable mid-IR emission in the PDR. \citet{Cesarsky2000b} reported the observation of amorphous silicates in Orion. More precisely, they detected their broad emission band, centred at 9.7~$\mu$m, in the whole \ion{H}{II} region and around the star $\theta^2$~Ori~A. \citet{Knight22} observed silicate emission in the \ion{H}{II} region only and found that this emission plummeted towards the IF of the Orion Bar. They proposed that the silicate grains were large and, while irradiation by Lyman-$\alpha$ photons in the \ion{H}{II} region would heat them sufficiently to cause significant emission at 9.7~$\mu$m, the PDR environment behind the IF would prevent such heating and emission. Our best model does not include isolated silicates but only silicates mixed with the population of carbonaceous grains in the form of large porous grains (the Mix(1:50) type grains presented in Sect. \ref{subsec:themis}). This has the effect of strongly attenuating the mid-IR vibrational bands of the silicates \citep[e.g.][]{Ysard2019}.
Then, above $\sim$13~$\mu$m, the shape of the model continuum does not match the observations either. This may be due to calibration issues (Channel 4 of MIRI/MRS) or differences in the equilibrium temperature of the large grains. This could also again be due to our simple modelling of aggregate grains based on spherical particles where materials are mixed according to effective medium theory \citep[see the description of Mix(1:50) type grains in][and Sect.~\ref{subsec:themis} for details]{Ysard2019}. However, calculating new aggregate optical properties would require a huge computational effort to explore the expected parameter space and is beyond the scope of this paper. 

In conclusion, it is too early to draw definitive conclusions regarding the observed differences between the model results and observations\footnote{In a forthcoming paper, we will explore these differences, once the data treatment and processing reach a higher level of maturity.}.
 
\section{Discussion}\label{sec:discussion}
\subsection{Nano-grain hydrogenation variation}

Our study provides key insights into the hydrogenation variation within the different regions of the Orion Bar. The analysis of the NIRSpec spectra shows variations in nano-grain properties from dense to irradiated regions. In comparison with the predicted THEMIS spectra for the DISM, we see that the continuum is underestimated in the \ion{H}{ii} region in front of the Bar, while the 3.4~$\mu$m band is significantly overestimated in the atomic region. However, the fit is reasonably good in the less excited molecular filament.

To address these discrepancies, we adjusted the grain properties in order to bring the continuum level within an acceptable range and adjust the band ratios, the grain band gap or hydrogenation level needs to be decreased. This suggests that the closer the region is to the star, the lower the hydrogen content in the grain structure\footnote{An inference that is consistent with the expected photoprocessing of a-C(:H) materials in intense radiation fields.}. Additionally, slightly larger grains (i.e. with higher minimum size) are found to be necessary in the irradiated regions, which is consistent with smaller grains being less resilient to harsh UV photons \citep[see for instance][and references therein]{schirmer2022}. 

The sensitivity of grain size and hydrogenation to the environment \citep{jones2012a,jones2012b,jones2012c} highlights the importance of considering these factors when analysing dust and gas observations in irradiated media. As shown for instance by \cite{schirmer2021}, the exact properties of the smallest hydrocarbons indeed impact the chemistry and dynamics in PDRs and thus our understanding of the location of the various ionization fronts\footnote{At the time of the \cite{schirmer2021} work, the strong spectroscopic constraints were not available and this significantly affected the data interpretation.}. The gas heating through photoelectric effect, mostly efficient for the smallest grains, as well as the H$_2$ molecule formation rate are also strongly dependent on $a_\text{min,a-C}$, $\alpha$, $M_\text{a-C}/M_\text{H}$, and $X_\text{H}$. By allowing us to compare the variations in the nano-grain properties for various $G_0/n_\text{H}$, the JWST observations will help us to improve our understanding of the stellar feedback influence on star formation, one of the main goals of the ERS program 1288.

\subsection{Comparisons with previous works}
\citet{schirmer2022} have proposed an explanation for observations of dust from mid- to far-IR wavelengths in the Orion Bar. Their study was based on photometric data from Spitzer IRAC and MIPS in the mid-IR, and from Herschel in the far-IR. 
Using the same methodology (DustEM, SOC, and THEMIS grain model), our results are comparable albeit with some distinctions. In their study, the degeneracy between $a_\text{min,a-C}$ and $\alpha$ did not allow them to draw a conclusion on the nano-grain minimum size. While \citet{schirmer2022} could only set an upper limit of $a_\text{min,a-C}$~$\leqslant$ 0.8~nm, our results suggest a smaller size of $a_\text{min,a-C}$~= 0.4~nm. Additionally, our model indicates a less steep grain size distribution ($\alpha$~= $-$5) compared to theirs ($\alpha$~$\leqslant$ $-$5.5).
They also found significant depletion of nano-grains (10$^{-5}$~$\leqslant$ $M_\text{a-C}/M_\text{H}$~$\leqslant$ 4~$\times$ 10$^{-5}$) compared to the gas and to their level in the diffuse ISM (0.17~$\times$ 10$^{-2}$). However, the abundance of nano-grains they found is four times lower than in our results $M_\text{a-C}/M_\text{H}$~= 1.09~$\times$ 10$^{-4}$.

We believe that the major difference explaining the discrepancies between our two models is the use of NIRSpec spectroscopic data. The spectroscopic data lead us to consider hydrogenation in the Orion Bar more than three times lower than in the diffuse medium, with the main constraint being the 3.3 to 3.4~$\mu$m band ratio. Insofar as \citet{schirmer2022} kept the value used for the diffuse medium, $E_\text{g}$~= 0.1~eV ($X_\text{H}$~$\sim$ 0.02), their model greatly overestimated the band at 3.4~$\mu$m (see middle panel in Fig.~\ref{fig:nirspec}). As the IRAC channel at 3.6~$\mu$m includes the two AIBs at 3.3 and 3.4~$\mu$m, the emission in this channel without any change in the size distribution is therefore also overestimated. In view of the free parameters of the \citet{schirmer2022} study, the only way to decrease the total flux in the IRAC channel at 3.6~$\mu$m is to increase $a_\text{min,a-C}$, the variation in the $\alpha$ parameter making it possible to keep a certain level of continuum in the IRAC channel at 4.5~$\mu$m.

The photo-dissociation of CH bonds by UV photons is known to be efficient in the ISM both from lab experiments and theory \citep[e.g.][]{Welch1972, Gruzdkov1994, Alata2014, Jones2014}. Variations in the 3.4 to 3.3~$\mu$m band ratios observed toward NGC 7023 also suggested variations in the hydrogen content of subnanometric carbonaceous particles \citep[or PAHs][]{pilleri_mixed_2015}. According to \citet{Jones2014}, the photo-processing timescale of small a-C:H particles would be only a few thousand years for the radiation field illuminating the Orion Bar. The use of JWST spectroscopic data allowed us to clearly demonstrate the efficiency of the hydrogen loss in such a PDR. It shows the value of the JWST and the PDRs4All program, in particular, for understanding stellar feedback in the interstellar medium. The scenario of equilibrium between the formation and destruction of the hydrocarbons responsible for emission in the AIBs will therefore have to be analysed in the light of these new results. However, this is beyond the scope of this paper as this will require a range of $G_0/n_{\rm H}$ to assess the efficiencies of all the physical and chemical processes involved. The necessary observations are currently being made for the GTO on the Horsehead ($G_0$~=100) and NGC 7023 ($G_0$~=1500) PDRs.

\subsection{Consequences of a 15-fold decrease in abundance of nano-grains on the PDR}

The nano-grain depletion found in this work is expected to have a profound influence on the physics and chemistry of the region, as discussed also in \cite{schirmer2021} in the case of the Horsehead nebula. Indeed in a PDR, the nano-grains directly control several key processes namely, the photoelectric gas heating, the UV extinction (see Fig.~\ref{fig:extuv}), and the grain surface where H$_2$ forms. As a result the temperature profile and H$_2$ emission from a PDR edge are expected to be affected when nano-grains are depleted. Preliminary simulations with an updated version of the Meudon PDR code where the dust model of this paper has been fully integrated and coupled to the gas, suggests that the depletion of nano-grains is reflected in the intensity ratio of rovibrational to pure rotational H$_2$ emission line (Meshaka et al., in prep.). The depletion in nano-grain abundance significantly influences the UV extinction curve, as clearly shown in Fig.~\ref{fig:extuv}. It is important to emphasize that this extinction curve is attributed to dust primarily situated in the atomic region of the Orion Bar, with a visual extinction in the UV illuminating direction (perpendicular to the Bar) $A_\text{V}$ $\lesssim$ 0.7 mag. This distinction is crucial as it underlines that the observed extinction is not a result of dust within the denser regions of the bar.

\section{Conclusion}\label{sec:conclusion}

We have investigated the dust emission by comparing model predictions with JWST/NIRCam and MIRI imaging, NIRSpec and MRS spectroscopic observations. We focussed on the illuminated edge of the Orion Bar, a highly irradiated dense PDR. We used radiative transfer modelling to investigate the spatial distribution of the dust emission along a cut through the Orion Bar PDR and compute the emerging spectrum. We firstly examined the spectra in the ionized, atomic, and warm molecular regions and then the spatial variations in 
all used filters at the PDR edge.
This scheme  is reliable, efficient, and allows us to probe  important variations of the dust properties (hydrogenation and abundance). The model incorporated essential parameters such as the density profile, incident radiation dilution factor, and the dust population consisting of THEMIS nano-grains transiently heated and larger pseudo-aggregates in thermal equilibrium with the radiation field. Our main results can be summarized as follows:

In the atomic region, the grains are characterized by a band gap energy of 0.03~eV, which is significantly smaller compared to the typical 0.1~eV in the diffuse ISM. This lower band gap energy indicates reduced hydrogenation levels in the grains, consistent with the NIRSpec observations of the Orion Bar. To our knowledge, this is the first time that hydrogenation variations can be traced through a PDR. For the molecular region, it was necessary to employ larger grains than in the diffuse ISM. We modelled profiles corresponding to the 3.3, 4.8 NIRCam, and 7.7, 11.3, 15, and 25 $\mu$m MIRI filters, subsequently checking the consistency with the MRS spectrum. 

Importantly, the investigation reveals that the nano-grains in the PDR edge are approximately 15 times less abundant than in the diffuse medium. This significant variation challenges the majority of existing PDR gas models that neglect such changes in nano-grain abundance. It highlights the importance of considering these variations in order to accurately estimate gas properties, including temperature, H$_2$ formation efficiency, and line intensities, which in turn will impact our understanding of the ongoing chemistry and dynamics in such a region. 

Due to the unprecedented high angular resolution of the JWST, the spatial distribution of the dust emission reveal a very sharp illuminated edge (on scales of $\sim$1$\arcsec$ or 0.002~pc) with a strong density gradient in the atomic region, just behind the IF. This strong density rise can be due to the sharp decrease in gas temperature at the IF, especially if the thermal pressures in the ionized and neutral regions are comparable. 

\begin{acknowledgements}
We are grateful to the anonymous referee for relevant and constructive comments.
MIRI data reduction is performed at the French MIRI centre of expertise with the support of CNES and the ANR-labcom INCLASS between IAS and the company ACRI-ST. NIRCam data reduction is performed between IRAP and IAS. This work is based on observations made with the NASA/ESA/CSA James Webb Space Telescope. The data were obtained from the Mikulski Archive for Space Telescopes at the Space Telescope Science Institute, which is operated by the Association of Universities for Research in Astronomy, Inc., under NASA contract NAS 5-03127 for JWST. These observations are associated with program \#1288.
Support for program \#1288 was provided by NASA through a grant from the Space Telescope Science Institute, which is operated by the Association of Universities for Research in Astronomy, Inc., under NASA contract NAS 5-03127.
Part of this work was supported by the Programme National ``Physique et Chimie du Milieu Interstellaire'' (PCMI) of CNRS/INSU with INC/INP co-funded by CEA and CNES. MEY aknowledges the financial support of CNES. MJ acknowledges the support of the Academy of Finland Grant No. 348342. MB acknowledges DST INSPIRE Faculty fellowship and Thanks The Inter-University Centre for Astronomy and Astrophysics for visiting associateship.
\end{acknowledgements}

\bibliographystyle{aa}
\bibliography{mybib}

\begin{thebibliography}{77}
\expandafter\ifx\csname natexlab\endcsname\relax\def\natexlab#1{#1}\fi

\bibitem[{{Alata} {et~al.}(2014){Alata}, {Cruz-Diaz}, {Mu{\~n}oz Caro}, \&
  {Dartois}}]{Alata2014}
{Alata}, I., {Cruz-Diaz}, G.~A., {Mu{\~n}oz Caro}, G.~M., \& {Dartois}, E.
  2014, \aap, 569, A119

\bibitem[{{Allamandola} {et~al.}(1985){Allamandola}, {Tielens}, \&
  {Barker}}]{allamandola_polycyclic_1985}
{Allamandola}, L.~J., {Tielens}, A.~G.~G.~M., \& {Barker}, J.~R. 1985, \apjl,
  290, L25

\bibitem[{{Arab} {et~al.}(2012){Arab}, {Abergel}, {Habart}, {Bernard-Salas},
  {Ayasso}, {Dassas}, {Martin}, \& {White}}]{arab2012}
{Arab}, H., {Abergel}, A., {Habart}, E., {et~al.} 2012, \aap, 541, A19

\bibitem[{{Bakes} \& {Tielens}(1994)}]{bakesandtielens94}
{Bakes}, E.~L.~O. \& {Tielens}, A.~G.~G.~M. 1994, \apj, 427, 822

\bibitem[{{Bernard-Salas} {et~al.}(2012){Bernard-Salas}, {Habart}, {Arab},
  {Abergel}, {Dartois}, {Martin}, {Bontemps}, {Joblin}, {White}, {Bernard}, \&
  {Naylor}}]{bernard-salas2012}
{Bernard-Salas}, J., {Habart}, E., {Arab}, H., {et~al.} 2012, \aap, 538, A37

\bibitem[{{Bern{\'e}} {et~al.}(2022{\natexlab{a}}){Bern{\'e}}, {Foschino},
  {Jalabert}, \& {Joblin}}]{berne2022}
{Bern{\'e}}, O., {Foschino}, S., {Jalabert}, F., \& {Joblin}, C.
  2022{\natexlab{a}}, \aap, 667, A159

\bibitem[{{Bern{\'e}} {et~al.}(2022{\natexlab{b}}){Bern{\'e}}, {Habart},
  {Peeters}, {Abergel}, {Bergin}, {Bernard-Salas}, {Bron}, {Cami}, {Dartois},
  {Fuente}, {Goicoechea}, {Gordon}, {Okada}, {Onaka}, {Robberto}, {R{\"o}llig},
  {Tielens}, {Vicente}, {Wolfire}, {Alarc{\'o}n}, {Boersma}, {Canin}, {Chown},
  {Dicken}, {Languignon}, {Le Gal}, {Pound}, {Trahin}, {Simmer}, {Sidhu}, {Van
  De Putte}, {Cuadrado}, {Guilloteau}, {Maragkoudakis}, {Schefter}, {Schirmer},
  {Cazaux}, {Aleman}, {Allamandola}, {Auchettl}, {Baratta}, {Bejaoui}, {Bera},
  {Bilalbegovi{\'c}}, {Black}, {Boulanger}, {Bouwman}, {Brandl}, {Brechignac},
  {Br{\"u}nken}, {Burkhardt}, {Candian}, {Cernicharo}, {Chabot}, {Chakraborty},
  {Champion}, {Colgan}, {Cooke}, {Coutens}, {Cox}, {Demyk}, {Donovan Meyer},
  {Engrand}, {Foschino}, {Garc{\'\i}a-Lario}, {Gavilan}, {Gerin}, {Godard},
  {Gottlieb}, {Guillard}, {Gusdorf}, {Hartigan}, {He}, {Herbst}, {Hornekaer},
  {J{\"a}ger}, {Janot-Pacheco}, {Joblin}, {Kaufman}, {Kemper}, {Kendrew},
  {Kirsanova}, {Klaassen}, {Knight}, {Kwok}, {Labiano}, {Lai}, {Lee},
  {Lefloch}, {Le Petit}, {Li}, {Linz}, {Mackie}, {Madden}, {Mascetti},
  {McGuire}, {Merino}, {Micelotta}, {Misselt}, {Morse}, {Mulas}, {Neelamkodan},
  {Ohsawa}, {Omont}, {Paladini}, {Palumbo}, {Pathak}, {Pendleton},
  {Petrignani}, {Pino}, {Puga}, {Rangwala}, {Rapacioli}, {Ricca},
  {Roman-Duval}, {Roser}, {Roueff}, {Rouill{\'e}}, {Salama}, {Sales},
  {Sandstrom}, {Sarre}, {Sciamma-O'Brien}, {Sellgren}, {Shannon}, {Shenoy},
  {Teyssier}, {Thomas}, {Togi}, {Verstraete}, {Witt}, {Wootten}, {Ysard},
  {Zettergren}, {Zhang}, {Zhang}, \& {Zhen}}]{berne2022ers}
{Bern{\'e}}, O., {Habart}, {\'E}., {Peeters}, E., {et~al.} 2022{\natexlab{b}},
  \pasp, 134, 054301

\bibitem[{{Bron} {et~al.}(2014){Bron}, {Le Bourlot}, \& {Le Petit}}]{bron14}
{Bron}, E., {Le Bourlot}, J., \& {Le Petit}, F. 2014, \aap, 569, A100

\bibitem[{{Carpentier} {et~al.}(2012){Carpentier}, {F{\'e}raud}, {Dartois},
  {Brunetto}, {Charon}, {Cao}, {d'Hendecourt}, {Br{\'e}chignac}, {Rouzaud}, \&
  {Pino}}]{carpentier2012}
{Carpentier}, Y., {F{\'e}raud}, G., {Dartois}, E., {et~al.} 2012, \aap, 548,
  A40

\bibitem[{Cesarsky {et~al.}(2000)Cesarsky, Jones, Lequeux, \&
  Verstraete}]{Cesarsky2000b}
Cesarsky, D., Jones, A.~P., Lequeux, J., \& Verstraete, L. 2000, \aap, 358, 708

\bibitem[{{Champion} {et~al.}(2017){Champion}, {Bern{\'e}}, {Vicente}, {Kamp},
  {Le Petit}, {Gusdorf}, {Joblin}, \& {Goicoechea}}]{Champion17}
{Champion}, J., {Bern{\'e}}, O., {Vicente}, S., {et~al.} 2017, \aap, 604, A69

\bibitem[{{Chastenet} {et~al.}(2021){Chastenet}, {Sandstrom}, {Chiang},
  {Hensley}, {Draine}, {Gordon}, {Koch}, {Leroy}, {Utomo}, \&
  {Williams}}]{Chastenet2021}
{Chastenet}, J., {Sandstrom}, K., {Chiang}, I.-D., {et~al.} 2021, \apj, 912,
  103

\bibitem[{{Chown} {et~al.}(2023){Chown}, {Sidhu}, {Peeters}, {Tielens}, {Cami},
  {Berne}, {Habart}, {Alarcon}, {Canin}, {Schroetter}, {Trahin}, {Van De
  Putte}, {Abergel}, {Bergin}, {Bernard-Salas}, {Boersma}, {Bron}, {Cuadrado},
  {Dartois}, {Dicken}, {El-Yajouri}, {Fuente}, {Goicoechea}, {Gordon}, {Issa},
  {Joblin}, {Kannavou}, {Khan}, {Lacinbala}, {Languignon}, {Le Gal},
  {Maragkoudakis}, {Meshaka}, {Okada}, {Onaka}, {Pasquini}, {Pound},
  {Robberto}, {Rollig}, {Schefter}, {Schirmer}, {Vicente}, {Wolfire},
  {Zannese}, {Aleman}, {Allamandola}, {Auchettl}, {Baratta}, {Bejaoui}, {Bera},
  {Black}, {Boulanger}, {Bouwman}, {Brandl}, {Brechignac}, {Brunken},
  {Buragohain}, {Burkhardt}, {Candian}, {Cazaux}, {Cernicharo}, {Chabot},
  {Chakraborty}, {Champion}, {Colgan}, {Cooke}, {Coutens}, {Cox}, {Demyk},
  {Donovan Meyer}, {Foschino}, {Garcia-Lario}, {Gavilan}, {Gerin}, {Gottlieb},
  {Guillard}, {Gusdorf}, {Hartigan}, {He}, {Herbst}, {Hornekaer}, {Jager},
  {Janot-Pacheco}, {Kaufman}, {Kemper}, {Kendrew}, {Kirsanova}, {Klaassen},
  {Kwok}, {Labiano}, {Lai}, {Lee}, {Lefloch}, {Le Petit}, {Li}, {Linz},
  {Mackie}, {Madden}, {Mascetti}, {McGuire}, {Merino}, {Micelotta}, {Misselt},
  {Morse}, {Mulas}, {Neelamkodan}, {Ohsawa}, {Omont}, {Paladini}, {Palumbo},
  {Pathak}, {Pendleton}, {Petrignani}, {Pino}, {Puga}, {Rangwala}, {Rapacioli},
  {Ricca}, {Roman-Duval}, {Roser}, {Roueff}, {Rouillee}, {Salama}, {Sales},
  {Sandstrom}, {Sarre}, {Sciamma-O'Brien}, {Sellgren}, {Shenoy}, {Teyssier},
  {Thomas}, {Togi}, {Verstraete}, {Witt}, {Wootten}, {Zettergren}, {Zhang},
  {Zhang}, \& {Zhen}}]{chown2023}
{Chown}, R., {Sidhu}, A., {Peeters}, E., {et~al.} 2023, arXiv e-prints,
  arXiv:2308.16733

\bibitem[{{Compi{\`e}gne} {et~al.}(2008){Compi{\`e}gne}, {Abergel},
  {Verstraete}, \& {Habart}}]{compiegne2008}
{Compi{\`e}gne}, M., {Abergel}, A., {Verstraete}, L., \& {Habart}, E. 2008,
  \aap, 491, 797

\bibitem[{{Compi{\`e}gne} {et~al.}(2011){Compi{\`e}gne}, {Verstraete}, {Jones},
  {Bernard}, {Boulanger}, {Flagey}, {Le Bourlot}, {Paradis}, \&
  {Ysard}}]{compiegne2011}
{Compi{\`e}gne}, M., {Verstraete}, L., {Jones}, A., {et~al.} 2011, \aap, 525,
  A103

\bibitem[{{Dartois} {et~al.}(2004){Dartois}, {Mu{\~n}oz Caro}, {Deboffle}, \&
  {d'Hendecourt}}]{dartois2004}
{Dartois}, E., {Mu{\~n}oz Caro}, G.~M., {Deboffle}, D., \& {d'Hendecourt}, L.
  2004, \aap, 423, L33

\bibitem[{{Duley}(1996)}]{duley1996}
{Duley}, W.~W. 1996, \mnras, 283, 343

\bibitem[{{Gorti} \& {Hollenbach}(2002)}]{Gorti02}
{Gorti}, U. \& {Hollenbach}, D. 2002, \apj, 573, 215

\bibitem[{Gruzdkov {et~al.}(1994)Gruzdkov, Watanabe, Sawabe, \&
  Matsumoto}]{Gruzdkov1994}
Gruzdkov, Y.~A., Watanabe, K., Sawabe, K., \& Matsumoto, Y. 1994, Chemical
  Physics Letters, 227, 243

\bibitem[{Guiu \& Bromley(2022)}]{Guiu2022}
Guiu, J.~M. \& Bromley, S.~T. 2022, {JPCA}, 126, 3854

\bibitem[{{Habart} {et~al.}(2023){Habart}, {Peeters}, {Bern{\'e}}, {Trahin},
  {Canin}, {Chown}, {Sidhu}, {Van De Putte}, {Alarc{\'o}n}, {Schroetter},
  {Dartois}, {Vicente}, {Abergel}, {Bergin}, {Bernard-Salas}, {Boersma},
  {Bron}, {Cami}, {Cuadrado}, {Dicken}, {Elyajouri}, {Fuente}, {Goicoechea},
  {Gordon}, {Issa}, {Joblin}, {Kannavou}, {Khan}, {Lacinbala}, {Languignon},
  {Le Gal}, {Maragkoudakis}, {Meshaka}, {Okada}, {Onaka}, {Pasquini}, {Pound},
  {Robberto}, {R{\"o}llig}, {Schefter}, {Schirmer}, {Tabone}, {Tielens},
  {Wolfire}, {Zannese}, {Ysard}, {Miville-Deschenes}, {Aleman}, {Allamandola},
  {Auchettl}, {Baratta}, {Bejaoui}, {Bera}, {Black}, {Boulanger}, {Bouwman},
  {Brandl}, {Brechignac}, {Br{\"u}nken}, {Buragohain}, {Burkhardt}, {Candian},
  {Cazaux}, {Cernicharo}, {Chabot}, {Chakraborty}, {Champion}, {Colgan},
  {Cooke}, {Coutens}, {Cox}, {Demyk}, {Donovan Meyer}, {Foschino},
  {Garc{\'\i}a-Lario}, {Gavilan}, {Gerin}, {Gottlieb}, {Guillard}, {Gusdorf},
  {Hartigan}, {He}, {Herbst}, {Hornekaer}, {J{\"a}ger}, {Janot-Pacheco},
  {Kaufman}, {Kemper}, {Kendrew}, {Kirsanova}, {Klaassen}, {Kwok}, {Labiano},
  {Lai}, {Lee}, {Lefloch}, {Le Petit}, {Li}, {Linz}, {Mackie}, {Madden},
  {Mascetti}, {McGuire}, {Merino}, {Micelotta}, {Misselt}, {Morse}, {Mulas},
  {Neelamkodan}, {Ohsawa}, {Omont}, {Paladini}, {Palumbo}, {Pathak},
  {Pendleton}, {Petrignani}, {Pino}, {Puga}, {Rangwala}, {Rapacioli}, {Ricca},
  {Roman-Duval}, {Roser}, {Roueff}, {Rouill{\'e}}, {Salama}, {Sales},
  {Sandstrom}, {Sarre}, {Sciamma-O'Brien}, {Sellgren}, {Shenoy}, {Teyssier},
  {Thomas}, {Togi}, {Verstraete}, {Witt}, {Wootten}, {Zettergren}, {Zhang},
  {Zhang}, \& {Zhen}}]{Habart2023jwst}
{Habart}, E., {Peeters}, E., {Bern{\'e}}, O., {et~al.} 2023, arXiv e-prints,
  arXiv:2308.16732

\bibitem[{{Habart} {et~al.}(2001){Habart}, {Verstraete}, {Boulanger}, {Pineau
  des For{\^e}ts}, {Le Peintre}, \& {Bernard}}]{Habart2001}
{Habart}, E., {Verstraete}, L., {Boulanger}, F., {et~al.} 2001, \aap, 373, 702

\bibitem[{{Habart} {et~al.}(2005){Habart}, {Walmsley}, {Verstraete}, {Cazaux},
  {Maiolino}, {Cox}, {Boulanger}, \& {Pineau des For{\^e}ts}}]{habart05}
{Habart}, E., {Walmsley}, M., {Verstraete}, L., {et~al.} 2005, \ssr, 119, 71

\bibitem[{{Haraguchi} {et~al.}(2012){Haraguchi}, {Nagayama}, {Kurita}, {Kino},
  \& {Sato}}]{Haraguchi2012}
{Haraguchi}, K., {Nagayama}, T., {Kurita}, M., {Kino}, M., \& {Sato}, S. 2012,
  \pasj, 64, 127

\bibitem[{{Henney}(2021)}]{Henney2021}
{Henney}, W.~J. 2021, \mnras, 502, 4597

\bibitem[{{Hensley} \& {Draine}(2021)}]{Hensley2021}
{Hensley}, B.~S. \& {Draine}, B.~T. 2021, \apj, 906, 73

\bibitem[{{Hogerheijde} {et~al.}(1995){Hogerheijde}, {Jansen}, \& {van
  Dishoeck}}]{Hoger95}
{Hogerheijde}, M.~R., {Jansen}, D.~J., \& {van Dishoeck}, E.~F. 1995, \aap,
  294, 792

\bibitem[{Hollenbach \& Tielens(1999)}]{hollenbach1999}
Hollenbach, D.~J. \& Tielens, A. G. G.~M. 1999, Reviews of Modern Physics, 71,
  173

\bibitem[{Iida {et~al.}(1985)Iida, Ohtaki, \& Seki}]{iida1985}
Iida, S., Ohtaki, T., \& Seki, T. 1985, in AIP Conference Proceedings, ed.
  P.~C. Taylor \& S.~G. Bishop, Vol. 120 (New York: AIP), 258

\bibitem[{{Jones}(2009)}]{jones2009}
{Jones}, A.~P. 2009, in Astronomical Society of the Pacific Conference Series,
  Vol. 414, Cosmic Dust - Near and Far, ed. T.~{Henning}, E.~{Gr{\"u}n}, \&
  J.~{Steinacker}, 473

\bibitem[{{Jones}(2012{\natexlab{a}})}]{jones2012a}
{Jones}, A.~P. 2012{\natexlab{a}}, \aap, 540, A1

\bibitem[{{Jones}(2012{\natexlab{b}})}]{jones2012b}
{Jones}, A.~P. 2012{\natexlab{b}}, \aap, 540, A2

\bibitem[{{Jones}(2012{\natexlab{c}})}]{jones2012c}
{Jones}, A.~P. 2012{\natexlab{c}}, \aap, 542, A98

\bibitem[{{Jones}(2014)}]{Jones2014}
{Jones}, A.~P. 2014, \planss, 100, 26

\bibitem[{{Jones}(2016)}]{jones2016c}
{Jones}, A.~P. 2016, Royal Society Open Science, 3, 160224

\bibitem[{{Jones} {et~al.}(2013){Jones}, {Fanciullo}, {K{\"o}hler},
  {Verstraete}, {Guillet}, {Bocchio}, \& {Ysard}}]{Jones2013}
{Jones}, A.~P., {Fanciullo}, L., {K{\"o}hler}, M., {et~al.} 2013, \aap, 558,
  A62

\bibitem[{{Jones} \& {Habart}(2015)}]{jones2015}
{Jones}, A.~P. \& {Habart}, E. 2015, \aap, 581, A92

\bibitem[{{Jones} {et~al.}(2017){Jones}, {K{\"o}hler}, {Ysard}, {Bocchio}, \&
  {Verstraete}}]{jones2017}
{Jones}, A.~P., {K{\"o}hler}, M., {Ysard}, N., {Bocchio}, M., \& {Verstraete},
  L. 2017, \aap, 602, A46

\bibitem[{{Juvela}(2019)}]{Juvela2019}
{Juvela}, M. 2019, \aap, 622, A79

\bibitem[{{Knight} {et~al.}(2022){Knight}, {Peeters}, {Tielens}, \&
  {Vacca}}]{Knight22}
{Knight}, C., {Peeters}, E., {Tielens}, A.~G.~G.~M., \& {Vacca}, W.~D. 2022,
  \mnras, 509, 3523

\bibitem[{{K{\"o}hler} {et~al.}(2014){K{\"o}hler}, {Habart}, {Arab},
  {Bernard-Salas}, {Ayasso}, {Abergel}, {Zavagno}, {Polehampton}, {van der
  Wiel}, {Naylor}, {Makiwa}, {Dassas}, {Joblin}, {Pilleri}, {Bern{\'e}},
  {Fuente}, {Gerin}, {Goicoechea}, \& {Teyssier}}]{kohler2014}
{K{\"o}hler}, M., {Habart}, E., {Arab}, H., {et~al.} 2014, \aap, 569, A109

\bibitem[{{K{\"o}hler} {et~al.}(2015){K{\"o}hler}, {Ysard}, \&
  {Jones}}]{kohler15}
{K{\"o}hler}, M., {Ysard}, N., \& {Jones}, A.~P. 2015, \aap, 579, A15

\bibitem[{Köhler {et~al.}(2012)Köhler, Stepnik, Jones, Guillet, Abergel,
  Ristorcelli, \& Bernard}]{kohler2012}
Köhler, M., Stepnik, B., Jones, A.~P., {et~al.} 2012, Astronomy and
  Astrophysics, 548, A61

\bibitem[{{Leger} \& {Puget}(1984)}]{leger_puget84}
{Leger}, A. \& {Puget}, J.~L. 1984, \aap, 500, 279

\bibitem[{{Marconi} {et~al.}(1998){Marconi}, {Testi}, {Natta}, \&
  {Walmsley}}]{marconi1998}
{Marconi}, A., {Testi}, L., {Natta}, A., \& {Walmsley}, C.~M. 1998, \aap, 330,
  696

\bibitem[{{Mathis} {et~al.}(1983){Mathis}, {Mezger}, \& {Panagia}}]{Mathis1983}
{Mathis}, J.~S., {Mezger}, P.~G., \& {Panagia}, N. 1983, \aap, 500, 259

\bibitem[{{Micelotta} {et~al.}(2010){Micelotta}, {Jones}, \&
  {Tielens}}]{micelotta2010}
{Micelotta}, E.~R., {Jones}, A.~P., \& {Tielens}, A.~G.~G.~M. 2010, \aap, 510,
  A36

\bibitem[{{Mori} {et~al.}(2014){Mori}, {Onaka}, {Sakon}, {Ishihara},
  {Shimonishi}, {Ohsawa}, \& {Bell}}]{mori2014}
{Mori}, T.~I., {Onaka}, T., {Sakon}, I., {et~al.} 2014, \apj, 784, 53

\bibitem[{{Mori} {et~al.}(2012){Mori}, {Sakon}, {Onaka}, {Kaneda}, {Umehata},
  \& {Ohsawa}}]{mori2012}
{Mori}, T.~I., {Sakon}, I., {Onaka}, T., {et~al.} 2012, \apj, 744, 68

\bibitem[{{O'Dell}(2001)}]{Odell01}
{O'Dell}, C.~R. 2001, \araa, 39, 99

\bibitem[{{Parvathi} {et~al.}(2012){Parvathi}, {Sofia}, {Murthy}, \&
  {Babu}}]{parvathi2012}
{Parvathi}, V.~S., {Sofia}, U.~J., {Murthy}, J., \& {Babu}, B.~R.~S. 2012,
  \apj, 760, 36

\bibitem[{{Peeters} {et~al.}(2023){Peeters}, {Habart}, {Berne}, {Sidhu},
  {Chown}, {Van De Putte}, {Trahin}, {Schroetter}, {Canin}, {Alarcon},
  {Schefter}, {Khan}, {Pasquini}, {Tielens}, {Wolfire}, {Dartois},
  {Goicoechea}, {Maragkoudakis}, {Onaka}, {Pound}, {Vicente}, {Abergel},
  {Bergin}, {Bernard-Salas}, {Boersma}, {Bron}, {Cami}, {Cuadrado}, {Dicken},
  {Elyajour}, {Fuente}, {Gordon}, {Issa}, {Joblin}, {Kannavou}, {Lacinbala},
  {Languignon}, {Le Gal}, {Meshaka}, {Okada}, {Robberto}, {Roellig},
  {Schirmer}, {Tabone}, {Zannese}, {Aleman}, {Allamandola}, {Auchettl},
  {Baratta}, {Bejaoui}, {Bera}, {Black}, {Boulanger}, {Bouwman}, {Brandl},
  {Brechignac}, {Brunken}, {Buragohain}, {Burkhardt}, {Candian}, {Cazaux},
  {Cernicharo}, {Chabot}, {Chakraborty}, {Champion}, {Colgan}, {Cooke},
  {Coutens}, {Cox}, {Demyk}, {Donovan Meyer}, {Foschino}, {Garcia-Lario},
  {Gerin}, {Gottlieb}, {Guillard}, {Gusdorf}, {Hartigan}, {He}, {Herbst},
  {Hornekaer}, {Jager}, {Janot-Pacheco}, {Kaufman}, {Kendrew}, {Kirsanova},
  {Klaassen}, {Kwok}, {Labiano}, {Lai}, {Lee}, {Lefloch}, {Le Petit}, {Li},
  {Linz}, {Mackie}, {Madden}, {Mascetti}, {McGuire}, {Merino}, {Micelotta},
  {Misselt}, {Morse}, {Mulas}, {Neelamkodan}, {Ohsawa}, {Paladini}, {Palumbo},
  {Pathak}, {Pendleton}, {Petrignani}, {Pino}, {Puga}, {Rangwala}, {Rapacioli},
  {Ricca}, {Roman-Duval}, {Roser}, {Roueff}, {Rouille}, {Salama}, {Sales},
  {Sandstrom}, {Sarre}, {Sciamma-O'Brien}, {Sellgren}, {Shenoy}, {Teyssier},
  {Thomas}, {Togi}, {Verstraete}, {Witt}, {Wootten}, {Ysard}, {Zettergren},
  {Zhang}, {Zhang}, \& {Zhen}}]{peeters2023arXiv}
{Peeters}, E., {Habart}, E., {Berne}, O., {et~al.} 2023, arXiv e-prints,
  arXiv:2310.08720

\bibitem[{Perrin {et~al.}(2014)Perrin, Sivaramakrishnan, Lajoie, Elliott,
  Pueyo, Ravindranath, \& Albert}]{Perrin20214}
Perrin, M.~D., Sivaramakrishnan, A., Lajoie, C.-P., {et~al.} 2014, in Space
  Telescopes and Instrumentation 2014: Optical, Infrared, and Millimeter Wave,
  ed. J.~M.~O. Jr., M.~Clampin, G.~G. Fazio, \& H.~A. MacEwen, Vol. 9143,
  International Society for Optics and Photonics (SPIE), 91433X

\bibitem[{{Pilleri} {et~al.}(2015){Pilleri}, {Joblin}, {Boulanger}, \&
  {Onaka}}]{pilleri2015}
{Pilleri}, P., {Joblin}, C., {Boulanger}, F., \& {Onaka}, T. 2015, \aap, 577,
  A16

\bibitem[{Pilleri {et~al.}(2015)Pilleri, Joblin, Boulanger, \&
  Onaka}]{pilleri_mixed_2015}
Pilleri, P., Joblin, C., Boulanger, F., \& Onaka, T. 2015, Astronomy \&
  Astrophysics, 577, A16

\bibitem[{{Pilleri} {et~al.}(2012){Pilleri}, {Montillaud}, {Bern{\'e}}, \&
  {Joblin}}]{pilleri2012}
{Pilleri}, P., {Montillaud}, J., {Bern{\'e}}, O., \& {Joblin}, C. 2012, A\&A,
  542, A69

\bibitem[{{Pino} {et~al.}(2008){Pino}, {Dartois}, {Cao}, {Carpentier},
  {Chamaill{\'e}}, {Vasquez}, {Jones}, {D'Hendecourt}, \&
  {Br{\'e}chignac}}]{pino2008}
{Pino}, T., {Dartois}, E., {Cao}, A.~T., {et~al.} 2008, \aap, 490, 665

\bibitem[{{Robertson} \& {O'Reilly}(1987)}]{robertson1987}
{Robertson}, J. \& {O'Reilly}, E.~P. 1987, \prb, 35, 2946

\bibitem[{{Savage} \& {Mathis}(1979)}]{Savage1979}
{Savage}, B.~D. \& {Mathis}, J.~S. 1979, \araa, 17, 73

\bibitem[{Schirmer {et~al.}(2020)Schirmer, Abergel, Verstraete, Ysard, Juvela,
  Jones, \& Habart}]{schirmer2020}
Schirmer, T., Abergel, A., Verstraete, L., {et~al.} 2020, Astronomy \&
  Astrophysics, 639, A144

\bibitem[{{Schirmer} {et~al.}(2021){Schirmer}, {Habart}, {Ysard}, {Bron}, {Le
  Bourlot}, {Verstraete}, {Abergel}, {Jones}, {Roueff}, \& {Le
  Petit}}]{schirmer2021}
{Schirmer}, T., {Habart}, E., {Ysard}, N., {et~al.} 2021, \aap, 649, A148

\bibitem[{{Schirmer} {et~al.}(2022){Schirmer}, {Ysard}, {Habart}, {Jones},
  {Abergel}, \& {Verstraete}}]{schirmer2022}
{Schirmer}, T., {Ysard}, N., {Habart}, E., {et~al.} 2022, \aap, 666, A49

\bibitem[{Sellgren {et~al.}(2010)Sellgren, Werner, Ingalls, Smith, Carleton, \&
  Joblin}]{sellgren2010c60}
Sellgren, K., Werner, M.~W., Ingalls, J.~G., {et~al.} 2010, The Astrophysical
  Journal Letters, 722, L54

\bibitem[{{Smith}(1984)}]{smith1984}
{Smith}, F.~W. 1984, Journal of Applied Physics, 55, 764

\bibitem[{{Sota} {et~al.}(2011){Sota}, {Ma{\'\i}z Apell{\'a}niz}, {Walborn},
  {Alfaro}, {Barb{\'a}}, {Morrell}, {Gamen}, \& {Arias}}]{sota2011}
{Sota}, A., {Ma{\'\i}z Apell{\'a}niz}, J., {Walborn}, N.~R., {et~al.} 2011,
  \apjs, 193, 24

\bibitem[{{Tamor} \& {Wu}(1990)}]{tamur1990}
{Tamor}, M.~A. \& {Wu}, C.~H. 1990, Journal of Applied Physics, 67, 1007

\bibitem[{{Tauber} {et~al.}(1995){Tauber}, {Lis}, {Keene}, {Schilke}, \&
  {Buettgenbach}}]{Tauber95}
{Tauber}, J.~A., {Lis}, D.~C., {Keene}, J., {Schilke}, P., \& {Buettgenbach},
  T.~H. 1995, \aap, 297, 567

\bibitem[{{Tielens} \& {Hollenbach}(1985)}]{Tielens_1985b}
{Tielens}, A.~G.~G.~M. \& {Hollenbach}, D. 1985, \apj, 291, 747

\bibitem[{{Tremblin} {et~al.}(2012){Tremblin}, {Audit}, {Minier}, {Schmidt}, \&
  {Schneider}}]{Tremblin2012}
{Tremblin}, P., {Audit}, E., {Minier}, V., {Schmidt}, W., \& {Schneider}, N.
  2012, \aap, 546, A33

\bibitem[{{Weingartner} \& {Draine}(2001)}]{weingartner_Draine_01_sizedistrib}
{Weingartner}, J.~C. \& {Draine}, B.~T. 2001, \apj, 548, 296

\bibitem[{{Welch} \& {Judge}(1972)}]{Welch1972}
{Welch}, A.~R. \& {Judge}, D.~L. 1972, \jcp, 57, 286

\bibitem[{Wen \& O'Dell(1995)}]{wen1995three}
Wen, Z. \& O'Dell, C. 1995, The Astrophysical Journal, 438, 784

\bibitem[{{Wen} \& {O'dell}(1995)}]{Wen1995}
{Wen}, Z. \& {O'dell}, C.~R. 1995, \apj, 438, 784

\bibitem[{Ysard {et~al.}(2013)Ysard, Abergel, Ristorcelli, Juvela, Pagani,
  Könyves, Spencer, White, \& Zavagno}]{ysard2013}
Ysard, N., Abergel, A., Ristorcelli, I., {et~al.} 2013, Astronomy and
  Astrophysics, 559, A133

\bibitem[{{Ysard} {et~al.}(2019){Ysard}, {Koehler}, {Jimenez-Serra}, {Jones},
  \& {Verstraete}}]{Ysard2019}
{Ysard}, N., {Koehler}, M., {Jimenez-Serra}, I., {Jones}, A.~P., \&
  {Verstraete}, L. 2019, \aap, 631, A88

\bibitem[{{Ysard} {et~al.}(2016){Ysard}, {K{\"o}hler}, {Jones}, {Dartois},
  {Godard}, \& {Gavilan}}]{ysard2016}
{Ysard}, N., {K{\"o}hler}, M., {Jones}, A., {et~al.} 2016, \aap, 588, A44

\bibitem[{{Zeegers} {et~al.}(2023){Zeegers}, {Guiu}, {Kemper}, {Marshall}, \&
  {Bromley}}]{Zeegers2023}
{Zeegers}, S.~T., {Guiu}, J.~M., {Kemper}, F., {Marshall}, J.~P., \& {Bromley},
  S.~T. 2023, Faraday Discussions, 245, 609

\end{thebibliography}
\begin{appendix} 
\section{Size distributions of dust in THEMIS}
\begin{figure*}
\centering
        \includegraphics[width=0.7\textwidth]{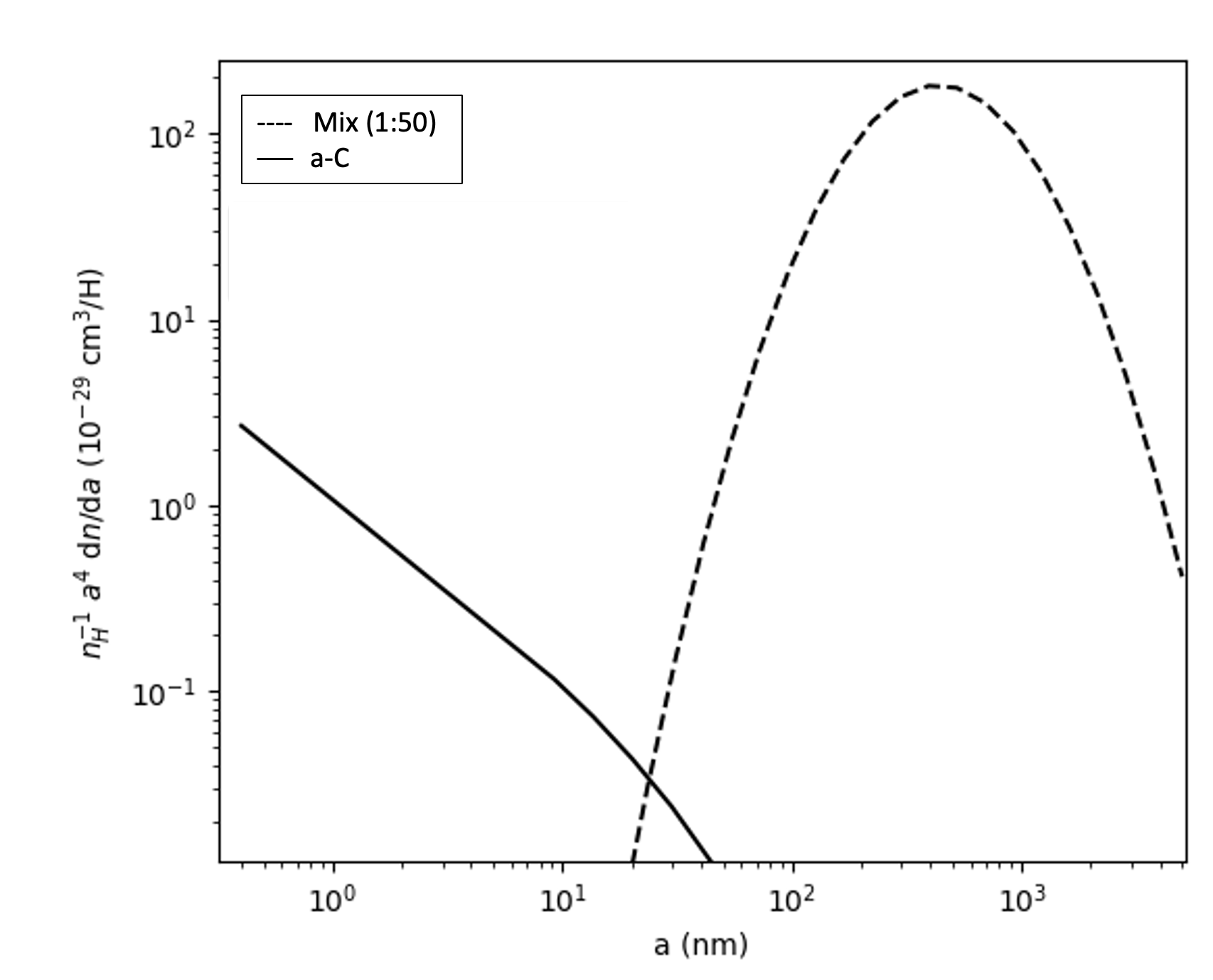}
    \caption{Size distributions of the dust mixtures for a-C grains from THEMIS (solid black line) and large grains \citep[i.e Mix 1:50 from][dash-dotted line]{Ysard2019}. Those are illustrative of our starting point since the parameters defining them are fitted (e.g. minimum and maximum sizes, slope of the power-law distributions, peak of the log-normal distributions). This figure is produced using DustEM \citep{compiegne2011}.}
    \label{fig:size_dist}
\end{figure*}
\FloatBarrier 

\section{UV extinction curve resulting from depleted nano grains and pseudo-agregrates}
\begin{figure*}
\centering
\includegraphics[width=0.7\textwidth]{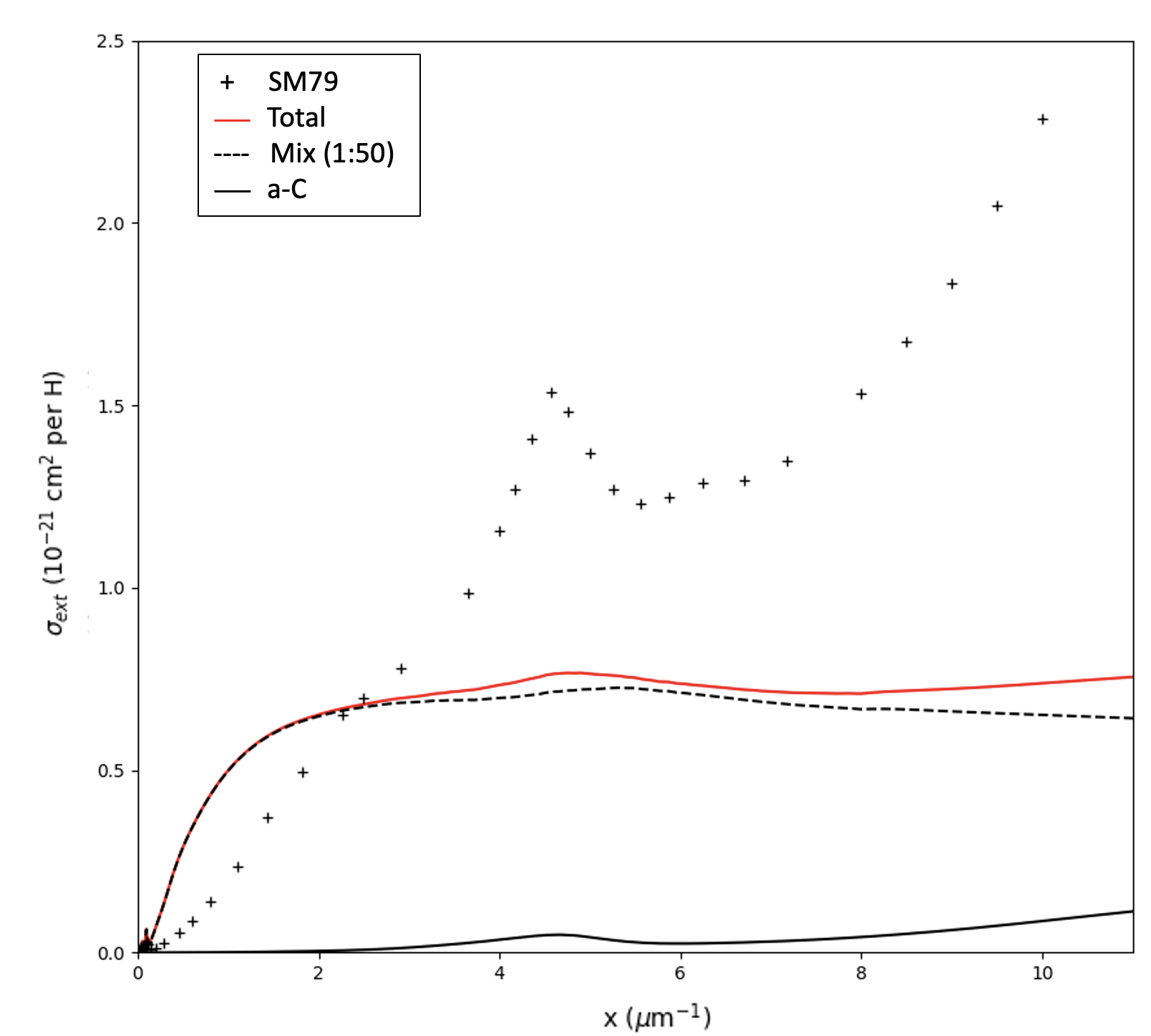}

\caption{UV extinction curve resulting from the grain distribution with the best parameters from our self-consistent model of the atomic region (red line). The contributions from the two grain distribution components are also shown in dashed black line for pseudo-aggregates and in solid black line for nano-grains a-C. For comparison, the standard extinction curve from \cite{Savage1979} (SM79) is represented by black cross symbols. The figure is produced using DustEM \citep{compiegne2011}.}
    \label{fig:extuv}
\end{figure*}
\FloatBarrier 

\section{Comparison of the observed (JWST) and modelled dust emission in the atomic PDR region of the Orion Bar PDR}
\FloatBarrier 
\begin{figure*}
    \centering
    \includegraphics[width=0.8\textwidth]{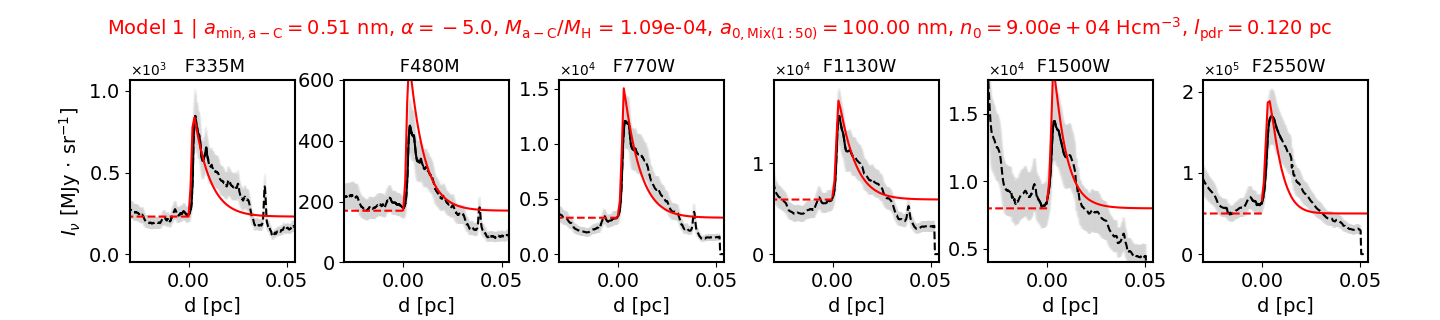}
    \includegraphics[width=0.8\textwidth]{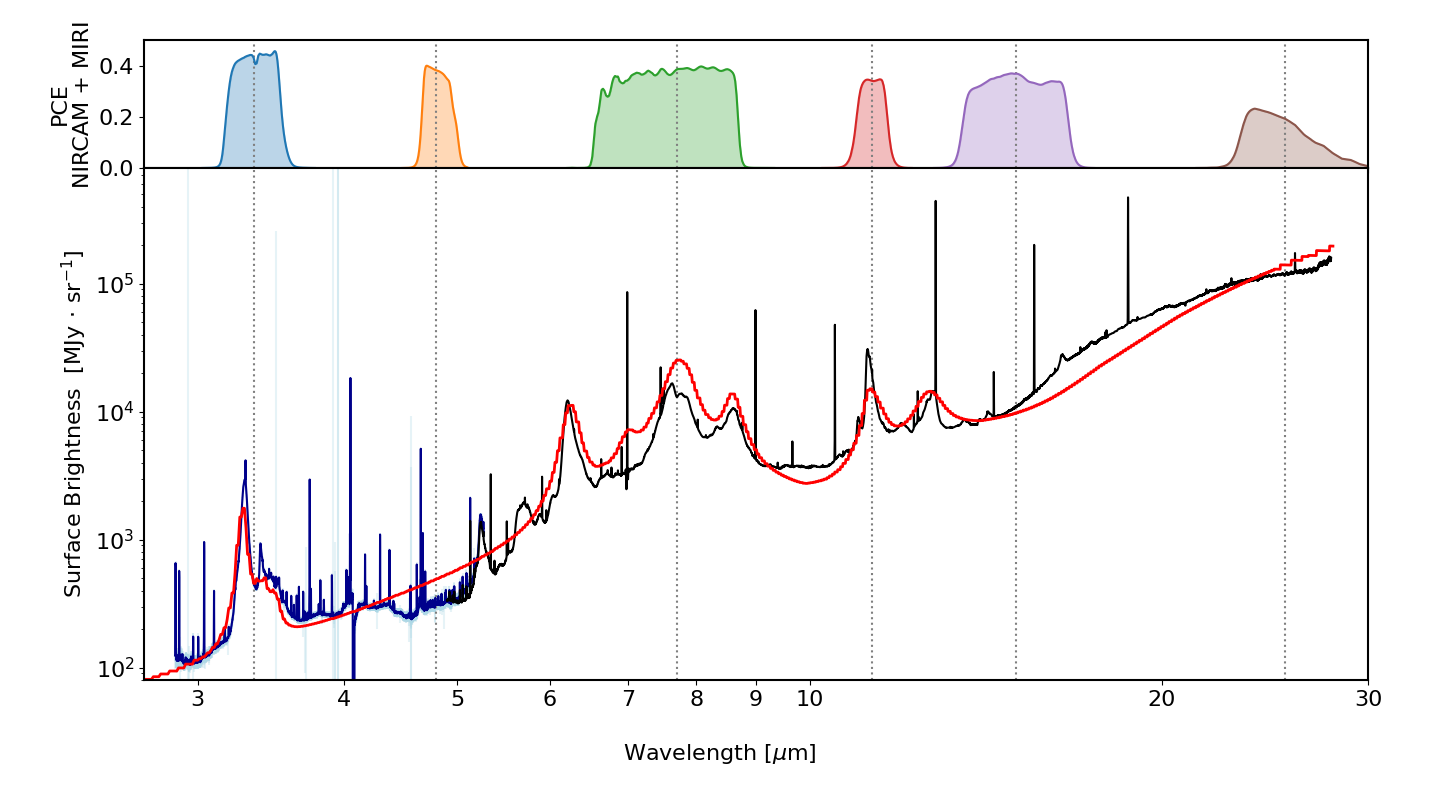}
    \caption{Same as Fig.~\ref{fig:bestmodels0} but for $a_\text{min,a-C}$~= 0.51~nm. This figure illustrates the sensitivity of the model to the parameters.}
    \label{fig:bestmodels2}
\end{figure*}

\end{appendix} 

\end{document}